\documentclass[%
 reprint,
 amsmath,amssymb,
 aps,
]{revtex4-2}

\usepackage{graphicx}
\usepackage{dcolumn}
\usepackage{bm}
\usepackage[justification=centering]{caption}
\usepackage{subcaption}
\usepackage{placeins}
\usepackage[export]{adjustbox}
\usepackage[ruled]{algorithm2e}
\SetKwInput{KwInput}{Input}               
\SetKwInput{KwOutput}{Output}   

\usepackage{tikz}
\usetikzlibrary{shapes,arrows}
\usetikzlibrary{decorations.pathmorphing}
\usetikzlibrary{decorations.pathreplacing}

\tikzset{
  font={\fontsize{12pt}{14}\selectfont}}
  
\usepackage{comment}

\usepackage{hyperref}

\usepackage{cleveref}

\begin{document}

\preprint{APS/123-QED}

\title{Microstructure synthesis using style-based generative adversarial network}

\author{Daria Fokina}
\email{daria.fokina@skoltech.ru}
\author{Ekaterina Muravleva}%
\author{George Ovchinnikov}
\author{Ivan Oseledets}
\affiliation{%
 Skolkovo Institute of Science and Technology
}%




\date{\today}

\begin{abstract}
Work considers the usage of StyleGAN architecture for the task of microstructure synthesis. The task is the following: given number of samples of structure we try to generate similar samples at the same time preserving its properties. Since the considered architecture is not able to produce samples of sizes larger than the training images, we propose to use image quilting to merge fixed-sized samples. One of the key features of the considered architecture is that it uses multiple image resolutions. We also investigate the necessity of such an approach. 
\end{abstract}

\keywords{Microstructure modelling, generative adversarial networks}
\maketitle


\section{Introduction}

\begin{figure*}
    \centering
    \scalebox{0.9}{
    \begin{subfigure}{0.3\textwidth}
    \centering
    \includegraphics[width=0.9\textwidth]{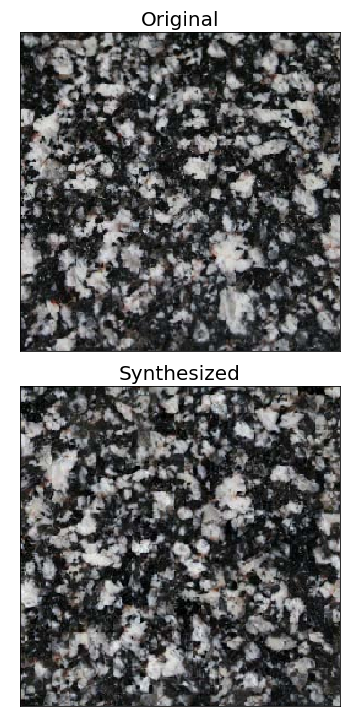}
    \caption{}
    \label{micro:dl_1}
    \end{subfigure}
    \begin{subfigure}{0.3\textwidth}
    \centering
    \includegraphics[width=0.9\textwidth]{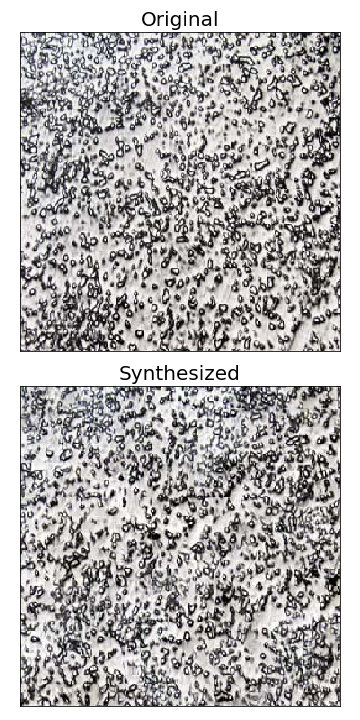}
    \caption{}
    \label{micro:dl_2}
    \end{subfigure}
    \begin{subfigure}{0.3\textwidth}
    \centering
    \includegraphics[width=0.9\textwidth]{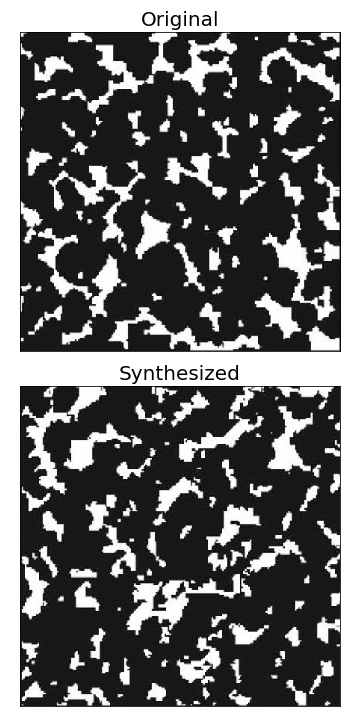}
    \caption{}
    \label{micro:ml}
    \end{subfigure}}
    \caption{ \label{micro:known}Examples of original and reconstructed images from \cite{low_dimesional_representation_nn} (Figs.~\ref{micro:dl_1}, \ref{micro:dl_2}) and \cite{ml_reconstruction} (Fig.~\ref{micro:ml})}
\end{figure*}

Recent papers have shown that generative models proposed in machine learning can be effectively used to synthesize complicated microstructures in different domain areas. 
Some examples of synthesized images are shown on Fig.~\ref{micro:known}. 
The generation process is represented as a mapping from a standard distribution (for example, $d$-dimensional normal distribution) to the sample space, in such a way that the generated structures are as close as possible to the real ones, which can be measured using different metrics. Learning such mappings is equivalent to learning the probabilistic generative process for the microstructures, and is useful in many cases, including generation of larger structures from several small samples, generation of microstructures with fixed properties and so on. 

Generative adversarial networks (GAN) are the main tool for generative modeling in deep learning and image processing, and since the introduction of GAN in \cite{goodfellow-gan}, tremendous progress has been achieved. One of the latest achievements is StyleGAN architecture \cite{stylegan}, which is able to generate surprisingly realistic faces. In this paper, we use this architecture to generate realistic-looking microstructures and obtain the results that are better than in the previous works that use GAN to generate microstructures.
One of the problems of StyleGAN is that it is only able to generate samples of the same size. We propose to generate larger images by generating patches and using quilting procedure proposed by Efros et al. \cite{quilting} by mixing the pixels on the boundary between patches. In this way, we can easily generate larger microstructures.

Main contributions of our paper are:
\begin{itemize}
    \item We propose to use StyleGAN architecture for microstructure synthesis.
    \item We propose to use image quilting between borders of two nearby patches to generate realistically looking samples of larger size.
    \item We test our method on two applications: microstructure synthesis and porous media reconstruction, and show that the generated structures are very close to the real ones in the effective properties. 
\end{itemize}

\section{Related Work}
One of the primary methods of reconstructing some structure from the given samples is the periodic unit cell (PUC) \cite{puc}. However, this method requires a proper cell construction, which can be rather complicated. Also, the resulting artificial structure is regular while the natural materials mostly have random heterogeneous organization. There exists a modification of a single cell repetition, which is called Wang Tilings \cite{wangtilings}, and it overcomes the main drawback of PUC. It includes the following steps: sampling several pieces of the original structure; combining them via quilting procedure \cite{quilting} to form a set of tiles; covering the plane with tiles randomly chosen from the prepared set. Such an approach allows generating irregular structures. 

Among other, more complex techniques, there are methods including computation of various spatial correlation functions \cite{cule1999generating, jiao2007modeling, jiao2008modeling, fullwood2008microstructure, jiao2009superior}. For this type of methods, the process is the following: to design a set of microstructure's descriptors and define an error between the given structure's descriptors and some generated realization's descriptors, then to find an appropriate realization via minimization of this error, for example by the simulated annealing procedure \cite{jiao2007modeling, jiao2008modeling}, or to use random sampling methods and generate structures until we reach an error value below the predefined threshold. More advanced models include neural networks \cite{cang2017microstructure, li2018transfer, low_dimesional_representation_nn}, in particular, generative adversarial networks (GANs).

Large deep learning models require a big amount of data and GAN models \cite{goodfellow-gan} were initially developed to create artificial data for training neural networks. Later GAN was used for material design \cite{materials_design_gan}. Authors of the paper aim to develop a model that will be able to create a microstructure by the given properties. For this purpose, first, a generative model is designed and trained, and then via Bayesian optimization, the desired structure is found. A GAN model was also applied to the microstructure reconstruction task. So, the suggested by Mosser et al. \cite{mosser} GAN approach is aimed at a quick generation of representative volume elements of the microstructure for the estimation of flow properties. However, further investigation \cite{case_study} shows that its result is not representative. In our work, we use another GAN architecture and achieve a better result for two-dimensional reconstructions.

\section{Generative modelling for microstructure synthesis}\label{stylegan}
\subsection{\label{problem_statement}Problem statement}
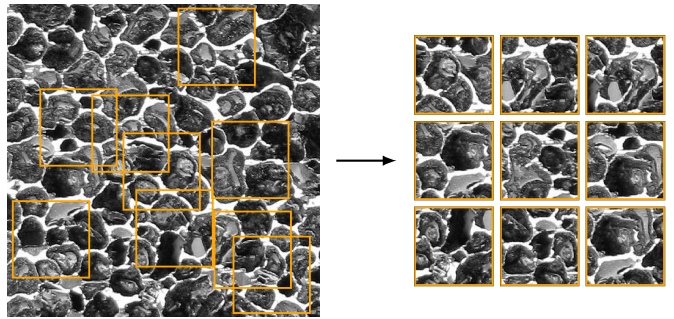
\begin{figure}
    \begin{tikzpicture}
    \node[inner sep=0pt] (train) at (0,0)
    {\includegraphics[width=.48\columnwidth]{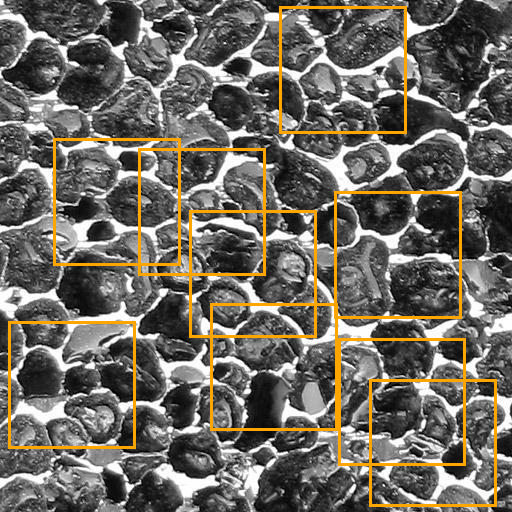}};
    \node[inner sep=0pt] (cuts) at (+5.0,0)
    {\includegraphics[width=.4\columnwidth]{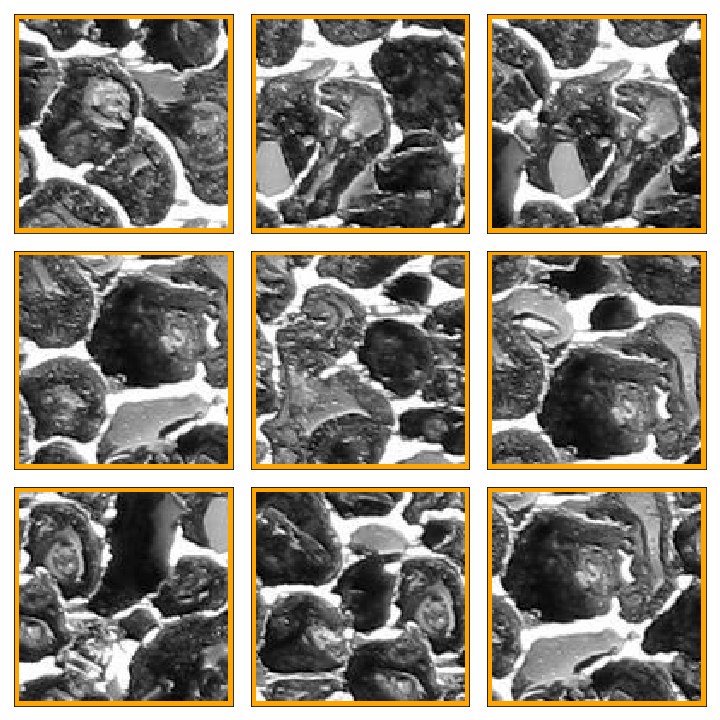}};
    \path (train.center) (+2.3,0) node(node1){};
    \path (train.center) (+3.1,0) node(node2){};
    \draw[-latex,thick] (node1.center) -- (node2.center);
    \end{tikzpicture}
    \caption{\label{micro:batch} An example of samples drawn from the given structure}
    
\end{figure}

To put the microstructure synthesis task into the framework of generative modeling, we do the following. Given a sample of the structure, we randomly take smaller pieces $x_1,\ldots,x_N$ from it (see Fig.~\ref{micro:batch}). We assume that $x_k$ are realisations of a random variable $x$ with the target probability distribution $p(x)$. We are willing to learn the mapping $G(z)$ such that if $z$ are sampled from a given distribution $p_z$ (typically, normal distribution), then $G(z)$ are distributed in the same way as $x$. In our case, we build $x_k$ as a subsample of a given microstructure sample by splitting it into smaller blocks, as shown on Fig.~\ref{micro:batch}. 

The mapping $G$ is called \emph{generator} of the generative model and is typically parameterized by a deep neural network architecture: $G(z) = G(z, \theta_G)$ \cite{goodfellow-gan, wgan, ls_gan}. In order to find $\theta_G$, we need to formulate the optimization problem for it. Let us introduce the function $\phi$, defining the similarity between original and generated distributions. Suppose $q$ is the distribution of generated samples $G(z,\theta_G)$. To find the optimal $\theta_D$, we need to solve the following minimization problem:
\begin{equation}\label{problem_statement:opt_problem}
    \min\limits_{\theta_G} \phi(p, q(\theta_G)).
\end{equation}

Among possible choices for $\phi$ are Total Variation, Kullback-Leibler divergence, Jensen-Shannon divergence, Earth Mover's Distance \cite{wgan}. We do not have the explicit formula for q and therefore cannot estimate $\theta_G$ directly. However, when $\phi$ represents Jensen-Shannon divergence, the problem is equivalent to the min-max problem:

\begin{equation}\label{problem_statement:jensen_shannon}
\begin{split}
JS(p||q) = \frac{1}{2} KL\left(p||\frac{1}{2}(p+q)\right) + \frac{1}{2} KL\left(q||\frac{1}{2}(p+q)\right) = \\
 = \arg\max\limits_D \frac{1}{2} \mathbb{E}_{x\sim p_\text{data}} \log D(x) + \frac{1}{2} \mathbb{E}_{z\sim p_z} \log (1- D(G(z))), 
\end{split}
\end{equation}
where $KL(p||q)$ is the Kullback-Leibler divergence:
\begin{equation}
    KL(p||q) = \mathbb{E}_{x\sim p} \log\frac{p(x)}{q(x)}.
\end{equation}

This problem formulation was introduced by Goodfellow et al. \cite{goodfellow-gan} and led to the concept of generative adversarial networks (GAN), where $D$ is represented by a parameterized mapping, called \emph{discriminator}. It maps generator outputs and real images to a real number in the interval [0,1]. Discriminator can be interpreted as a classifier, whose aim is to distinguish between real and fake samples. Its output is the predicted probability that the input sample is real. The generator tries to produce realizations close to the original and fool the discriminator. 
Denote the output of generator as $G(z,\theta_G)$, for discriminator - $D(x, \theta_D)$. In this case the optimization problem is formulated as:
\begin{equation}\label{problem_statement:gan_loss}
\begin{split}
    \min\limits_{\theta_G}\max\limits_{\theta_D} ( &\mathbb{E}_{x\sim p_{\text{data}}} \log D(x, \theta_D) +\\ & + \mathbb{E}_{z\sim p_z}\log(1-D(G(z,\theta_G), \theta_D)).
\end{split}
\end{equation}
In computations the expected values are replaced with the empirical mean, and this problem is solved via a stochastic gradient-type method. During one step of the method parameters of the generator are fixed and parameters of the discriminator are optimized, and then, vice versa, parameters of the discriminator are fixed and parameters of the generator are optimized (Algorithm \ref{problem_statement:algo}, \cite{goodfellow-gan}). 

\begin{algorithm}
\caption{\label{problem_statement:algo}GAN Training}

\SetAlgoLined
\KwInput{$N$ \,---\, number of iterations, $k_G$ \,---\, number of generator updates, $k_D$ \,---\, number of discriminator updates, $\alpha$ \,---\, learning rate, $k$ \,---\, batch size}
\KwOutput{}
\For{$i = 1,\ldots,N$}{
\For{$j = 1,\ldots,k_D$}{
sample $x_1,\ldots,x_k$ from real samples\\
sample $z_1,\ldots,z_k$ from noise distribution\\
\begin{equation*}
\begin{split}
    \theta_D \leftarrow -\alpha\nabla_{\theta_D} (& \frac{1}{k} \sum\limits_{l=1}^k \log D(x_l,\theta_D)+\\ & + \frac{1}{k} \sum \limits_{l=1}^k \log D(G(z_l, \theta_G),\theta_D))   
\end{split}
\end{equation*}
}
\For{$j = 1,\ldots,k_G$}{
sample $z_1,\ldots,z_k$ from noise distribution\\
\begin{equation*}
\theta_G \leftarrow -\alpha\nabla_{\theta_G} \left(\frac{1}{k} \sum \limits_{l=1}^k \log D(G(z_l, \theta_G),\theta_D)\right)    
\end{equation*}
}
}
\end{algorithm}

\subsection{Wasserstein GAN}
Another popular choice for the metrics $\phi$ introduced in Section \ref{problem_statement} is Earth Mover's Distance:
\begin{equation}
    \phi(p,q) = W(p, q) = \inf \limits_{\gamma\in \prod(p,q)} \mathbb{E}_{(x,y)\sim \prod(p,q)} \|x-y\|,
\end{equation}
where $p,\, q$ are two given distributions, $\prod(p,q)$ is a distribution from a set of distributions, whose marginals equal $p$ and $q$ respectively. 
Let us consider a compact metric space $(\chi, d)$, $p, q$ - probability measures on $\chi$ and a mapping $f: \chi \rightarrow \mathbb{R}$. According to the Kantorovich-Rubinstein duality theorem \cite{kantorovich_rubinstein_theorem}, $W(p,q)$ can also be rewritten as:
\begin{equation}
\begin{split}
    W(p, q) & =  \sup\limits_{\|f\|_\text{Lip} \le 1} \left\{\int\limits_\chi f(x)dp(x) - \int\limits_\chi f(y)dq(y) \right\} =\\
    &  =  \sup\limits_{\|f\|_\text{Lip} \le 1} \{\mathbb{E}_{x\sim p} f(x) - \mathbb{E}_{x\sim q} f(x)\},
\end{split}
\end{equation}
where 
\begin{equation*}
    \|f\|_\text{Lip} = \inf \left\{C\in \mathbb{R} \Big| |f(x) - f(y)| \le Cd(x,y) \forall\, x,y \in \chi \right\}.
\end{equation*} 
With this choice of the distance between distributions we get the Wasserstein GAN and the minmax problem is reformulated as follows:
\begin{equation}
    \min\limits_{\theta_G}\max\limits_{\theta_D} \mathbb{E}_{z\sim p_z} D(G(z,\theta_G), \theta_D) -\mathbb{E}_{x\sim p_{\text{data}}} D(x, \theta_D).
\end{equation}

The new formulation requires the mapping $D$ to be $1$-Lipschitz. Instead of 1-Lipschitz requirement, often \emph{gradient penalty} is introduced \cite{wgan_gp}:
\begin{equation}
    \text{GP} (\hat{x}) = \|\nabla_{\theta_D} D(\hat{x}, \theta_D)-1\|_2^2, 
\end{equation}
where $\hat{x} = \epsilon G(z) + (1-\epsilon) x$, $x \sim p,\, z\sim p_z,\, \epsilon\sim U([0,1])$. The gradient penalty is added to the optimized function and the final problem is stated as:
\begin{equation}
\begin{split}
     \min\limits_{\theta_G}\max\limits_{\theta_D} \mathbb{E}_{z\sim p_z} D(G(z,\theta_G), \theta_D) -\mathbb{E}_{x\sim p_{\text{data}}} D(x, \theta_D) +\\+ \lambda \|\nabla_{\theta_D} D(\hat{x})-1\|_2^2.
\end{split}
\end{equation}

\subsection{Basic concepts of neural networks}
To represent $G$ and $D$ we will use deep convolutional neural networks, since they are very efficient in image processing tasks (e.g. image style transfer \cite{style_transfer}, super-resolution \cite{superresolution}, semantic segmentation \cite{segmentation}, image generation \cite{wgan}).
For the completeness of the description, here we define what the neural network is and what are its constituents. A more comprehensive introduction can be found in the book by Goodfellow et al. \cite{Goodfellow-et-al-2016}
\paragraph{Fully-connected network.}
Neural networks are usually represented as a sequence of \emph{layers}, and each layer consists of \emph{neurons}. The primary type of a neural network is a fully-connected network (FC network). Each of its neurons maps an input vector $x$ to a scalar via transformation $f(Ax+b)$, $Ax+b$ is a number, $f$ is a non-linear function and is called an \emph{activation} function. Combination of outputs of neurons in the layer form a new vector input for the next layer, and so the vector is forwarded (Fig.~\ref{fc_network}). Usually, the input of a network is a collection, or a \emph{batch}, of vectors. In order to find the best parameters $A$ and $b$ for each neuron the output of the network is then forwarded to the loss function, and the parameters are found through the loss minimization, normally via stochastic gradient descent or its modifications, e.g., Adam optimizer \cite{adam}. 
\pgfdeclarelayer{background}
\pgfdeclarelayer{foreground}
\pgfsetlayers{background,main,foreground}

\tikzstyle{arrow_style} = [-latex, draw, ultra thin]
\tikzstyle{neuron}=[draw, circle, minimum size=0.5cm, fill=white]
\tikzstyle{boldfont} = [minimum height=5em, text centered]

\begin{figure}[h]
\centering
\captionsetup{justification=centering}
    \begin{tikzpicture}
    \node (input1) [neuron] {};
    \path (input1.center)+(0,+0.7) node (input2) [neuron] {};
    \path (input1.center)+(0,-0.7) node (input3) [neuron] {};
    \path (input3.center)+(0,-0.9) node (input) {input};
    
    \path (input1.center)+(+2,+1.1) node (n1) [neuron] {};
    \path (n1.center)+(0,-0.7) node (n2) [neuron] {};
    \path (n2.center)+(0,-0.7) node (n3) [neuron] {};
    \path (n3.center)+(0,-0.7) node (n4) [neuron] {};
    
    \draw [arrow_style] (input1.east) -- (n1.west);
    \draw [arrow_style] (input1.east) -- (n2.west);
    \draw [arrow_style] (input1.east) -- (n3.west);
    \draw [arrow_style] (input1.east) -- (n4.west);
    
    \draw [arrow_style] (input2.east) -- (n1.west);
    \draw [arrow_style] (input2.east) -- (n2.west);
    \draw [arrow_style] (input2.east) -- (n3.west);
    \draw [arrow_style] (input2.east) -- (n4.west);
    
    \draw [arrow_style] (input3.east) -- (n1.west);
    \draw [arrow_style] (input3.east) -- (n2.west);
    \draw [arrow_style] (input3.east) -- (n3.west);
    \draw [arrow_style] (input3.east) -- (n4.west);
    
    \path(n4.center)+(0,-1) node (l1) [align=center] {layer 1};
    
    \path (n1.center)+(+2,-0.4) node(n5) [neuron] {};
    \path (n5.center)+(0,-0.7) node(n6) [neuron] {};
    \path (n6.center)+(0,-0.7) node(n7) [neuron] {};
    
    \path (l1.center)+(+2,0) node (l2) [align=center] {layer 2};
    
    \draw [arrow_style] (n1.east) -- (n5.west);
    \draw [arrow_style] (n1.east) -- (n6.west);
    \draw [arrow_style] (n1.east) -- (n7.west);
    
    \draw [arrow_style] (n2.east) -- (n5.west);
    \draw [arrow_style] (n2.east) -- (n6.west);
    \draw [arrow_style] (n2.east) -- (n7.west);
    
    \draw [arrow_style] (n3.east) -- (n5.west);
    \draw [arrow_style] (n3.east) -- (n6.west);
    \draw [arrow_style] (n3.east) -- (n7.west);
    
    \draw [arrow_style] (n4.east) -- (n5.west);
    \draw [arrow_style] (n4.east) -- (n6.west);
    \draw [arrow_style] (n4.east) -- (n7.west);
    
    \path (n6.center)+(+2,+0.35) node (output1) [neuron] {};
    \path (output1.center)+(0,-0.7) node(output2) [neuron] {};
    \path (output2.center)+(0,-0.9) node (output) {output};
    
    \draw [arrow_style] (n5.east) -- (output1.west);
    \draw [arrow_style] (n6.east) -- (output1.west);
    \draw [arrow_style] (n7.east) -- (output1.west);
    
    \draw [arrow_style] (n5.east) -- (output2.west);
    \draw [arrow_style] (n6.east) -- (output2.west);
    \draw [arrow_style] (n7.east) -- (output2.west);

    \begin{pgfonlayer}{background}
            \path (input2.north east)+(+0.2,+0.2) node (a) {};
            \path (input3.south west)+(-0.2,-0.2) node (b) {};
            \path[fill=orange!10,rounded corners, draw=black]
                (a) rectangle (b);
                
            \path (output1.north east)+(+0.2,+0.2) node (a) {};
            \path (output2.south west)+(-0.2,-0.2) node (b) {};
            \path[fill=orange!10,rounded corners, draw=black]
                (a) rectangle (b);
                
            \path (n1.north east)+(+0.2,+0.2) node (a) {};
            \path (n4.south west)+(-0.2,-0.2) node (b) {};
            \path[fill=green!10,rounded corners, draw=black]
                (a) rectangle (b);
                
            \path (n5.north east)+(+0.2,+0.2) node (a) {};
            \path (n7.south west)+(-0.2,-0.2) node (b) {};
            \path[fill=green!10,rounded corners, draw=black]
                (a) rectangle (b);
    \end{pgfonlayer}
    \end{tikzpicture}
\caption{A 3-layer fully-connected neural network (input is usually not regarded as a layer)}
\label{fc_network}
\end{figure}
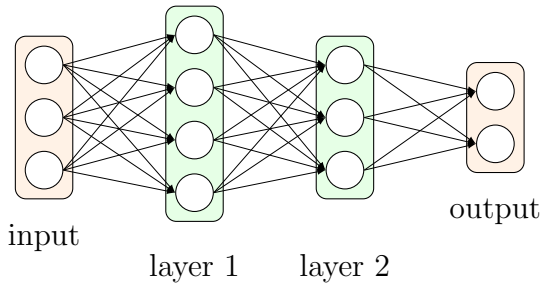
\paragraph{Convolutional network.}
Another type of neural network is a convolutional neural network. In this case the term ``neuron'' is replaced by the notion of a filter and the input is usually represented by an image. Here we assume that all the operations have a two-dimensional input, but they can be easily extended to 1D and 3D cases. The convolutional layer is represented as a number of filters. Each filter performs an operation of discrete convolution with each channel of the input and sums the result channel-wise. The parameters of this network are the values of filters, and they are found using the same stochastic optimization procedure. Like FC network each convolution in this type of network is typically followed by a non-linearity.

There exists an extension to a simple convolution. It is named deconvolution \cite{deconv}, or fractionally-strided convolution, and is used to increase the size of the input. In fact, it is the same operation of convolution, but at first, a zero-padding to each pixel is added (Fig.~ \ref{stylegan:deconv}).

\pgfdeclarelayer{background}
\pgfdeclarelayer{foreground}
\pgfsetlayers{background,main,foreground}

\begin{figure*}
    \centering
    \captionsetup{justification=centering}
    
\hspace{-50pt}    
\begin{subfigure}{0.15\textwidth}
\scalebox{0.5}{  
    \begin{tikzpicture}
    \begin{pgfonlayer}{foreground}
    \path [draw, black, fill=blue!30] (1.3,1.1) -- (1.3,3.5) -- (2.8,4.4) -- (2.8,2) -- cycle;
    \path [draw, black] (1.3, 1.9) -- (2.8, 2.8);
    \path [draw, black] (1.3, 2.7) -- (2.8, 3.6);
    \path [draw, black] (1.8, 1.4) -- (1.8, 3.8);
    \path [draw, black] (2.3, 1.7) -- (2.3, 4.1);
    \end{pgfonlayer}
    \begin{pgfonlayer}{main}
    \path [draw, dashed, black] (-2,-1.2) -- (-2,4.4) -- (1.5,6.5) -- (1.5,0.9)  -- cycle;
    
    \path [draw, black, fill = green!40] (-1,1) -- (-0.5,1.3) -- (-0.5,2.1) -- (-1,1.8)  -- cycle;
    
    \path [draw, black, fill = green!40] (0,1.6) -- (0.5,1.9) -- (0.5,2.7) -- (0,2.4)  -- cycle;
    
    \path [draw, black, fill = green!40] (0, 3.2) -- (0.5,3.5) -- (0.5,4.3) -- (0, 4)  -- cycle;
    
    \path [draw, black, fill = green!40] (-1,2.6) -- (-0.5,2.9) -- (-0.5,3.7) -- (-1,3.4)  -- cycle;
    
    \path [draw, dashed, black] (-1.5, -0.9) -- (-1.5, 4.7);
    \path [draw, dashed, black] (-1, -0.6) -- (-1, 5);
    \path [draw, dashed, black] (-0.5, -0.3) -- (-0.5, 5.3);
    \path [draw, dashed, black] (0, 0) -- (0, 5.6);
    \path [draw, dashed, black] (0.5, 0.3) -- (0.5, 5.9);
    \path [draw, dashed, black] (1, 0.6) -- (1, 6.2);
    
    \path [draw, dashed, black] (-2, -0.4) -- (1.5, 1.7);
    \path [draw, dashed, black] (-2, 0.4) -- (1.5, 2.5);
    \path [draw, dashed, black] (-2, 1.2) -- (1.5, 3.3);
    \path [draw, dashed, black] (-2, 2) -- (1.5, 4.1);
    \path [draw, dashed, black] (-2, 2.8) -- (1.5, 4.9);
    \path [draw, dashed, black] (-2, 3.6) -- (1.5, 5.7);

    \path [draw, black, thick] (-2,4.4) -- (1.3,3.5);
    \path [draw, black, thick] (-2,2) -- (1.3,1.1);
    \path [draw, black, thick] (-0.5,2.9) -- (2.8,2);
    \path [draw, black, thick] (-0.5,5.3) -- (2.8,4.4);

    \path [draw, black, thick] (-2,4.4) -- (-2,2) -- (-0.5,2.9) -- (-0.5,5.3) -- cycle;
    
    \end{pgfonlayer}
    \end{tikzpicture}
}
\end{subfigure}
\hspace{5pt}
\begin{subfigure}{0.15\textwidth}
\scalebox{0.5}{  
    \begin{tikzpicture}
    \begin{pgfonlayer}{foreground}
    \path [draw, black, fill=blue!30] (1.3,1.1) -- (1.3,3.5) -- (2.8,4.4) -- (2.8,2) -- cycle;
    \path [draw, black] (1.3, 1.9) -- (2.8, 2.8);
    \path [draw, black] (1.3, 2.7) -- (2.8, 3.6);
    \path [draw, black] (1.8, 1.4) -- (1.8, 3.8);
    \path [draw, black] (2.3, 1.7) -- (2.3, 4.1);
    \end{pgfonlayer}
    \begin{pgfonlayer}{main}
    \path [draw, dashed, black] (-2,-1.2) -- (-2,4.4) -- (1.5,6.5) -- (1.5,0.9)  -- cycle;
    
    \path [draw, black, fill = green!40] (-1,1) -- (-0.5,1.3) -- (-0.5,2.1) -- (-1,1.8)  -- cycle;
    
    \path [draw, black, fill = green!40] (0,1.6) -- (0.5,1.9) -- (0.5,2.7) -- (0,2.4)  -- cycle;
    
    \path [draw, black, fill = green!40] (0, 3.2) -- (0.5,3.5) -- (0.5,4.3) -- (0, 4)  -- cycle;
    
    \path [draw, black, fill = green!40] (-1,2.6) -- (-0.5,2.9) -- (-0.5,3.7) -- (-1,3.4)  -- cycle;
    
    \path [draw, dashed, black] (-1.5, -0.9) -- (-1.5, 4.7);
    \path [draw, dashed, black] (-1, -0.6) -- (-1, 5);
    \path [draw, dashed, black] (-0.5, -0.3) -- (-0.5, 5.3);
    \path [draw, dashed, black] (0, 0) -- (0, 5.6);
    \path [draw, dashed, black] (0.5, 0.3) -- (0.5, 5.9);
    \path [draw, dashed, black] (1, 0.6) -- (1, 6.2);
    
    \path [draw, dashed, black] (-2, -0.4) -- (1.5, 1.7);
    \path [draw, dashed, black] (-2, 0.4) -- (1.5, 2.5);
    \path [draw, dashed, black] (-2, 1.2) -- (1.5, 3.3);
    \path [draw, dashed, black] (-2, 2) -- (1.5, 4.1);
    \path [draw, dashed, black] (-2, 2.8) -- (1.5, 4.9);
    \path [draw, dashed, black] (-2, 3.6) -- (1.5, 5.7);

    \path [draw, black, thick] (-1.5,4.7) -- (1.3,3.5);
    \path [draw, black, thick] (-1.5,2.3) -- (1.3,1.1);
    \path [draw, black, thick] (0,3.2) -- (2.8,2);
    \path [draw, black, thick] (0,5.6) -- (2.8,4.4);

    \path [draw, black, thick] (-1.5,4.7) -- (-1.5,2.3) -- (0,3.2) -- (0,5.6) -- cycle;
    \path [draw, -latex, black, thick] (-5,2) -- (-3, 2);

    \end{pgfonlayer}
    \end{tikzpicture}
}
\end{subfigure}
\hspace{45pt}
\begin{subfigure}{0.15\textwidth}
\scalebox{0.5}{  
    \begin{tikzpicture}
    \begin{pgfonlayer}{foreground}
    \path [draw, black, fill=blue!30] (1.3,1.1) -- (1.3,3.5) -- (2.8,4.4) -- (2.8,2) -- cycle;
    \path [draw, black] (1.3, 1.9) -- (2.8, 2.8);
    \path [draw, black] (1.3, 2.7) -- (2.8, 3.6);
    \path [draw, black] (1.8, 1.4) -- (1.8, 3.8);
    \path [draw, black] (2.3, 1.7) -- (2.3, 4.1);
    \end{pgfonlayer}
    \begin{pgfonlayer}{main}
    \path [draw, dashed, black] (-2,-1.2) -- (-2,4.4) -- (1.5,6.5) -- (1.5,0.9)  -- cycle;
    
    \path [draw, black, fill = green!40] (-1,1) -- (-0.5,1.3) -- (-0.5,2.1) -- (-1,1.8)  -- cycle;
    
    \path [draw, black, fill = green!40] (0,1.6) -- (0.5,1.9) -- (0.5,2.7) -- (0,2.4)  -- cycle;
    
    \path [draw, black, fill = green!40] (0, 3.2) -- (0.5,3.5) -- (0.5,4.3) -- (0, 4)  -- cycle;
    
    \path [draw, black, fill = green!40] (-1,2.6) -- (-0.5,2.9) -- (-0.5,3.7) -- (-1,3.4)  -- cycle;
    
    \path [draw, dashed, black] (-1.5, -0.9) -- (-1.5, 4.7);
    \path [draw, dashed, black] (-1, -0.6) -- (-1, 5);
    \path [draw, dashed, black] (-0.5, -0.3) -- (-0.5, 5.3);
    \path [draw, dashed, black] (0, 0) -- (0, 5.6);
    \path [draw, dashed, black] (0.5, 0.3) -- (0.5, 5.9);
    \path [draw, dashed, black] (1, 0.6) -- (1, 6.2);
    
    \path [draw, dashed, black] (-2, -0.4) -- (1.5, 1.7);
    \path [draw, dashed, black] (-2, 0.4) -- (1.5, 2.5);
    \path [draw, dashed, black] (-2, 1.2) -- (1.5, 3.3);
    \path [draw, dashed, black] (-2, 2) -- (1.5, 4.1);
    \path [draw, dashed, black] (-2, 2.8) -- (1.5, 4.9);
    \path [draw, dashed, black] (-2, 3.6) -- (1.5, 5.7);

    \path [draw, black, thick] (-1,5) -- (1.3,3.5);
    \path [draw, black, thick] (-1,2.6) -- (1.3,1.1);
    \path [draw, black, thick] (0.5,3.5) -- (2.8,2);
    \path [draw, black, thick] (0.5,5.9) -- (2.8,4.4);

    \path [draw, black, thick] (-1,5) -- (-1,2.6) -- (0.5,3.5) -- (0.5,5.9) -- cycle;
    \path [draw, -latex, black, thick] (-5,2) -- (-3, 2);
    \end{pgfonlayer}
    \end{tikzpicture}
}
\end{subfigure}
    \caption{\label{stylegan:deconv} Deconvolution with stride 2 and kernel $3\times3$, green pixels - pixels of the image, white is padding, blue - kernel of convolution}
\end{figure*}
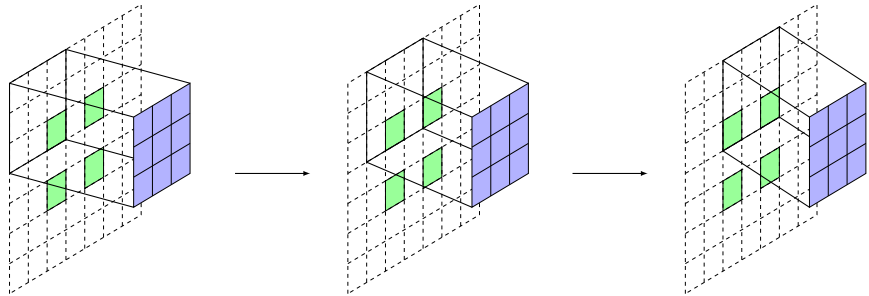

Due to the huge number of parameters, neural networks usually tend to overfit the input data easily. In order to prevent it, different types of regularization are used. For example, we can add \emph{pooling} layers to the network after convolution. This layer slides with a window along the input and outputs one number from the values observed in the window. This procedure decreases the number of network parameters and hence prevents overfitting. Typically used operations in the window are maximum and averaging operations (max and average pooling, correspondingly).

Usual convolutional neural networks observe only given fixed size samples. However, for microstructures description multiscale methods are widely used \cite{multiscale_crack_model, multiscale_heterogeneous_layers, multiscale_solidification}. There are multiple ways to use different scales in a neural network. One of them is to use different image resolutions for different stages of training. This idea was initially suggested for generating faces by Karras et al. \cite{pggan}. The generator architecture was later modified and presented as a style-based generator \cite{stylegan}, and the new architecture (named StyleGAN) shows impressive results. Now we proceed with a more detailed description of a StyleGAN architecture.

\subsection{StyleGAN architecture.}
Like any GAN, StyleGAN consists of a generator and a discriminator. Both networks can be represented as sequences of upsampling and downsampling blocks correspondingly. Detailed schemes can be viewed on Figs.~\ref{stylegan:generator}, \ref{stylegan:discriminator}.

\pgfdeclarelayer{background}
\pgfdeclarelayer{foreground}
\pgfsetlayers{background,main,foreground}

\tikzstyle{arrow_style} = [-latex, draw, thick]
\tikzstyle{discr}=[draw,  minimum height=3em, text width=14em, text centered, rounded corners, fill=white]
\tikzstyle{noframe} = [minimum height=3em, text width=14em, text centered]

\tikzstyle{gen}=[draw,  minimum height=3em, text width=14em, text centered, rounded corners, fill=white]
\tikzstyle{noframe1} = [minimum height=3em, text width=3em, text centered]
\tikzstyle{noise} = [draw, rounded corners, minimum height=3em, fill=white]
\tikzstyle{ab} = [draw, minimum height=1em, text width=1em, text centered, fill=white]

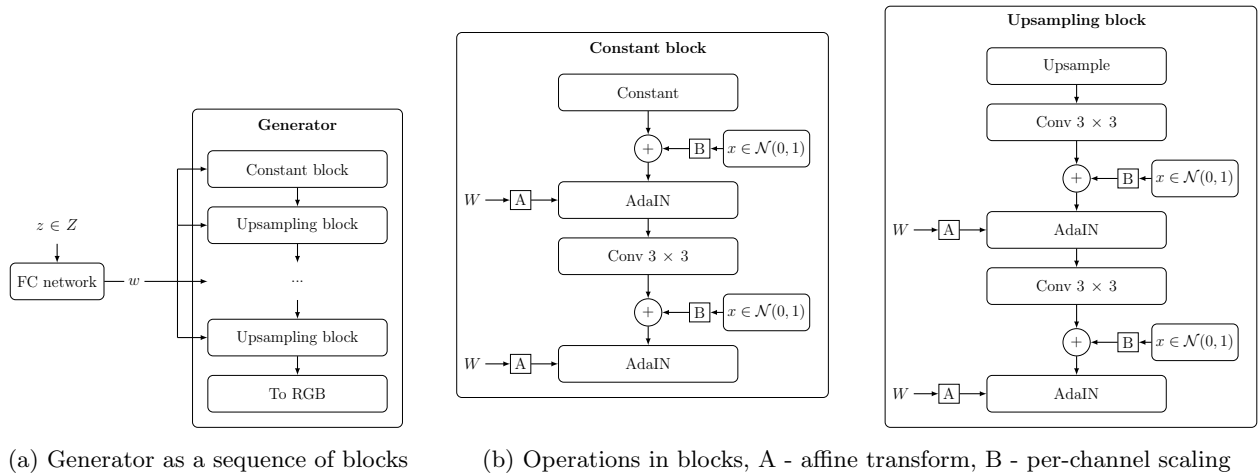
\begin{figure*}
\begin{subfigure}[t]{0.3\textwidth}
\centering
\captionsetup{justification=centering}
\scalebox{0.5}{
    \begin{tikzpicture}
    \node (const) [gen]{Constant block};
    \path (const.north)+(0,+0.7) node(gen)[noframe] {\large{\textbf{Generator}}};
    \path (const.south)+(0,-1) node(ups1)[gen] {Upsampling block};
    \draw [arrow_style] (const.south) -- (ups1.north);
    \path (ups1.south)+(0,-1) node(empty)[noframe] {...};
    \draw [arrow_style] (ups1.south) -- (empty.north);
    \path (empty.south)+(0,-1) node(ups2)[gen] {Upsampling block};
    \draw [arrow_style] (empty.south) -- (ups2.north);
    \path (ups2.south)+(0,-1) node(torgb)[gen] {To RGB};
    \draw [arrow_style] (ups2.south) -- (torgb.north);
    \path (empty.west)+(-4,0) node(fc)[draw,  minimum height=3em, text width=7em, text centered, rounded corners, fill=white] {FC network};
    \path (fc.east)+(+0.8,0) node(w){$w$};
    \draw [thick] (fc.east) -- (w);
    \path (ups1.west)+(-0.8,0) node(con1){};
    \draw [arrow_style] (con1.center) -- (ups1.west);
    \path (ups2.west)+(-0.8,0) node(con2){};
    \draw [arrow_style] (con2.center) -- (ups2.west);
    \path (empty.west)+(-0.8,0) node(con3){};
    \draw [thick] (con1.center) -- (con3.center);
    \draw [thick] (con2.center) -- (con3.center);
    \draw [thick] (w) -- (con3.center);
    \draw [arrow_style] (con3.center) --(empty.west);
    \path (const.west)+(-0.8,0) node(con4){};
    \draw [arrow_style] (con4.center) -- (const.west);
    \draw [thick] (con4.center) -- (con1.center);
    
    \path (fc.north) + (0,+1) node(z)[minimum height=3em, text width=5em, text centered]{$z\in Z$};
    \draw[arrow_style] (z.south) -- (fc.north);
    \begin{pgfonlayer}{background}
            \path (const.north east)+(+0.4,1.1) node (a) {};
            \path (torgb.south west)+(-0.4,-0.4) node (b) {};
            \path[fill=white!10,rounded corners, draw=black]
                (a) rectangle (b);
    \end{pgfonlayer}
    \end{tikzpicture}
}
\caption{\label{generator:seq}Generator as a sequence of blocks}
\end{subfigure}
\begin{subfigure}[t]{0.65\textwidth}
\centering
\captionsetup{justification=centering}
\scalebox{0.5}{
    \begin{tikzpicture}
    \node (const)[gen]{Constant};
    \path (const.north)+(0,+0.7) node(constblock)[noframe] {\large{\textbf{Constant block}}};
    \path (const.south)+(0,-1) node(plus1)[circle, draw, fill=white] {+};
    \draw [arrow_style] (const.south) -- (plus1.north);
    \path (plus1.east)+(+1,0) node(b1)[ab] {B};
    \draw [arrow_style] (b1.west) -- (plus1.east);
    \path (plus1.south)+(0,-1) node(adain1)[gen]{AdaIN};
    \draw [arrow_style] (plus1.south) -- (adain1.north);
    \path (adain1.west)+(-1,0) node(a1)[ab] {A};
    \draw [arrow_style] (a1.east) -- (adain1.west);
    \path (a1.west)+(-1,0) node(w1) {$W$};
    \draw [arrow_style] (w1.east) -- (a1.west);
    \path (adain1.south)+(0,-1) node(conv)[gen]{Conv $3\times3$};
    \draw [arrow_style] (adain1.south) -- (conv.north);
    \path (conv.south)+(0,-1) node(plus2)[circle, draw, fill=white] {+};
    \draw [arrow_style] (conv.south) -- (plus2.north);
    \path (plus2.east)+(+1,0) node(b2)[ab] {B};
    \draw [arrow_style] (b2.west) -- (plus2.east);
    \path (plus2.south)+(0,-1) node(adain2)[gen]{AdaIN};
    \draw [arrow_style] (plus2.south) -- (adain2.north);
    \path (adain2.west)+(-1,0) node(a2)[ab] {A};
    \draw [arrow_style] (a2.east) -- (adain2.west);
    \path (a2.west)+(-1,0) node(w2) {$W$};
    \draw [arrow_style] (w2.east) -- (a2.west);
    \path (b2.east)+(+1.5,0) node(noise2)[noise]{$x\in \mathcal{N}(0,1)$};
    \draw [arrow_style] (noise2.west) -- (b2.east);
    \path (b1.east)+(+1.5,0) node(noise1)[noise]{$x\in \mathcal{N}(0,1)$};
    \draw [arrow_style] (noise1.west) -- (b1.east);
    
    \path (const.east)+(+9,+0.7) node(ups)[gen]{Upsample};
    \path (ups.north)+(0,+0.7) node(upsblock)[noframe] {\large{\textbf{Upsampling block}}};
    \path (ups.south) + (0,-1) node(conv1)[gen]{Conv $3\times3$};
    \draw[arrow_style] (ups.south) -- (conv1.north);
    \path (conv1.south)+(0,-1) node(plus3)[circle, draw, fill=white] {+};
    \draw [arrow_style] (conv1.south) -- (plus3.north);
    \path (plus3.east)+(+1,0) node(b3)[ab] {B};
    \draw [arrow_style] (b3.west) -- (plus3.east);
    \path (plus3.south)+(0,-1) node(adain3)[gen]{AdaIN};
    \draw [arrow_style] (plus3.south) -- (adain3.north);
    \path (adain3.west)+(-1,0) node(a3)[ab] {A};
    \draw [arrow_style] (a3.east) -- (adain3.west);
    \path (a3.west)+(-1,0) node(w3) {$W$};
    \draw [arrow_style] (w3.east) -- (a3.west);
    \path (adain3.south)+(0,-1) node(conv2)[gen]{Conv $3\times3$};
    \draw [arrow_style] (adain3.south) -- (conv2.north);
    \path (conv2.south)+(0,-1) node(plus4)[circle, draw, fill=white] {+};
    \draw [arrow_style] (conv2.south) -- (plus4.north);
    \path (plus4.east)+(+1,0) node(b4)[ab] {B};
    \draw [arrow_style] (b4.west) -- (plus4.east);
    \path (plus4.south)+(0,-1) node(adain4)[gen]{AdaIN};
    \draw [arrow_style] (plus4.south) -- (adain4.north);
    \path (adain4.west)+(-1,0) node(a4)[ab] {A};
    \draw [arrow_style] (a4.east) -- (adain4.west);
    \path (a4.west)+(-1,0) node(w4) {$W$};
    \draw [arrow_style] (w4.east) -- (a4.west);
    \path (b4.east)+(+1.5,0) node(noise4)[noise]{$x\in \mathcal{N}(0,1)$};
    \draw [arrow_style] (noise4.west) -- (b4.east);
    \path (b3.east)+(+1.5,0) node(noise3)[noise]{$x\in \mathcal{N}(0,1)$};
    \draw [arrow_style] (noise3.west) -- (b3.east);

    \begin{pgfonlayer}{background}
            \path (const.north east)+(+2.4,+1.1) node (a) {};
            \path (adain2.south west)+(-2.7,-0.4) node (b) {};
            \path[fill=white!10,rounded corners, draw=black]
                (a) rectangle (b);
            
            \path (ups.north east)+(+2.4,1.1) node (a) {};
            \path (adain4.south west)+(-2.7,-0.4) node (b) {};
            \path[fill=white!10,rounded corners, draw=black]
                (a) rectangle (b);

    \end{pgfonlayer}
    \end{tikzpicture}

}
\caption{Operations in blocks, A - affine transform, B - per-channel scaling}
\label{generator:operations}
\end{subfigure}
\caption{\label{stylegan:generator}Generator scheme}
\end{figure*}

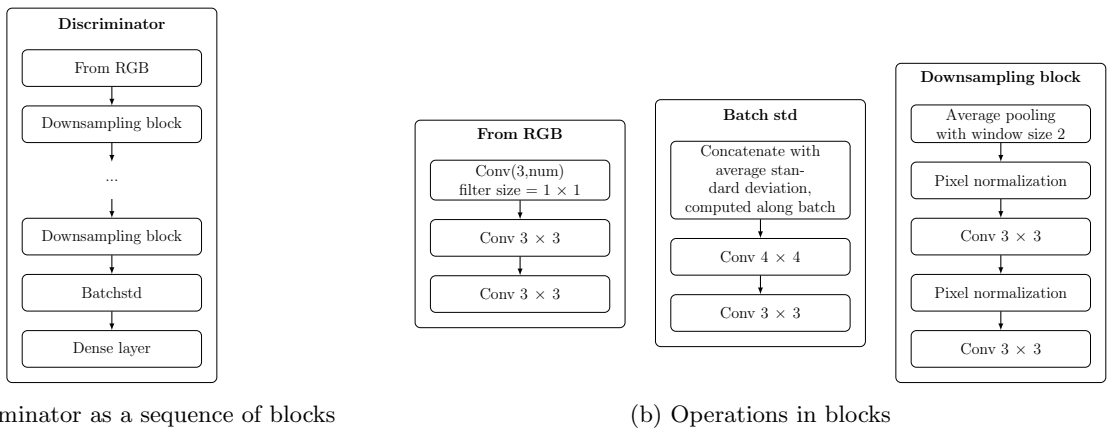
\begin{figure*}
\centering
\begin{subfigure}[t]{0.35\textwidth}
\centering
\captionsetup{justification=centering}
\scalebox{0.5}{
    \begin{tikzpicture}
    \node (fromrgb)[discr, fill=white!30]{From RGB};
    \path (fromrgb.north)+(0,+0.7) node(Discr)[noframe] {\large{\textbf{Discriminator}}};
    \path (fromrgb.south)+(0,-1) node(ds1)[discr, fill=white!30] {Downsampling block};
    \draw [arrow_style] (fromrgb.south) -- (ds1.north);
    \path (ds1.south)+(0,-1) node(empty)[noframe] {...};
    \draw [arrow_style] (ds1.south) -- (empty.north);
    \path (empty.south)+(0,-1) node(ds2)[discr, fill=white!30] {Downsampling block};
    \draw [arrow_style] (empty.south) -- (ds2.north);
    \path (ds2.south)+(0,-1) node(bstd)[discr, fill=white!30] {Batchstd};
    \draw [arrow_style] (ds2.south) -- (bstd.north);
    \path (bstd.south)+(0,-1) node(dense)[discr, fill=white!30] {Dense layer};
    \draw [arrow_style] (bstd.south) -- (dense.north);
    \begin{pgfonlayer}{background}
            \path (fromrgb.north east)+(+0.4,1.1) node (a) {};
            \path (dense.south west)+(-0.4,-0.4) node (b) {};
            \path[fill=white!10,rounded corners, draw=black]
                (a) rectangle (b);
    \end{pgfonlayer}
    \end{tikzpicture}
}
\caption{\label{discriminator:seq} Discriminator as a sequence of blocks}
\end{subfigure}
\begin{subfigure}[t]{0.6\textwidth}
\centering
\captionsetup{justification=centering}
\scalebox{0.5}{
    \begin{tikzpicture}
    \node (conv1)[discr, fill=white!30]{Conv(3,num)\\filter size = $1\times1$};
    \path (conv1.north)+(0,+0.7) node(Discr)[noframe] {\large{\textbf{From RGB}}};
    \path (conv1.south)+(0,-1) node(conv2)[discr, fill=white!30] {Conv $3\times3$};
    \draw [arrow_style] (conv1.south) -- (conv2.north);
    \path (conv2.south)+(0,-1) node(conv3)[discr, fill=white] {Conv $3\times3$};
    \draw [arrow_style] (conv2.south) -- (conv3.north);
    
    \path (conv1.east)+(+4,0) node(concat)[discr, fill=white]{Concatenate with \\average standard deviation, \\computed along batch};
    \path (concat.north)+(0,+0.7) node(bstd)[noframe] {\large{\textbf{Batch std}}};
    \path (concat.south)+(0,-1) node(conv4)[discr, fill=white] {Conv $4\times4$};
    \draw [arrow_style] (concat.south) -- (conv4.north);
    \path (conv4.south)+(0,-1) node(conv5)[discr, fill=white] {Conv $3\times3$};
    \draw [arrow_style] (conv4.south) -- (conv5.north);
    
    \path (concat.east)+(+4,+1.5)
    node(avgpool)[discr, fill=white] {Average pooling with window size 2};
    \path (avgpool.north)+(0,+0.7) node(dsblock)[noframe]{\large{\textbf{Downsampling block}}};
    \path (avgpool.south)+(0,-1) node(pn1)[discr] {Pixel normalization};
    \draw [arrow_style] (avgpool.south) -- (pn1.north);
    \path (pn1.south)+(0,-1) node(conv6)[discr] {Conv $3\times3$};
    \draw [arrow_style] (pn1.south) -- (conv6.north);
    \path (conv6.south)+(0,-1) node(pn2)[discr] {Pixel normalization};
    \draw [arrow_style] (conv6.south) -- (pn2.north);
    \path (pn2.south)+(0,-1) node(conv7)[discr] {Conv $3\times3$};
    \draw [arrow_style] (pn2.south) -- (conv7.north);

    \begin{pgfonlayer}{background}
            \path (fromrgb.north east)+(+0.4,1.1) node (a) {};
            \path (conv3.south west)+(-0.4,-0.4) node (b) {};
            \path[fill=white!10,rounded corners, draw=black]
                (a) rectangle (b);
                
            \path (concat.north east)+(+0.4,1.1) node (a) {};
            \path (conv5.south west)+(-0.4,-0.4) node (b) {};
            \path[fill=white!10,rounded corners, draw=black]
                (a) rectangle (b);

            \path (avgpool.north east)+(+0.4,1.1) node (a) {};
            \path (conv7.south west)+(-0.4,-0.4) node (b) {};
            \path[fill=white!10,rounded corners, draw=black]
                (a) rectangle (b);
    \end{pgfonlayer}
    \end{tikzpicture}
}
\caption{Operations in blocks}
\label{discriminator:operations}

\end{subfigure}
\caption{\label{stylegan:discriminator} Discriminator scheme}
\end{figure*}

As one can see on Fig.~\ref{stylegan:generator}, the generator network starts with constant, which has to be learned. This constant is then forwarded through several upsampling blocks (Fig.~\ref{generator:operations}). Operation \emph{Upsample} is performed by the deconvolution layer. A block, transforming the output to an image, finishes the chain. 
There are several introduced important components:
\begin{itemize}
    \item summation of outputs with scaled per-channel Gaussian noise between deconvolutions - accounts for random features (for faces it was freckles, hair location and other)
    \item more sophisticated normalization layer, called adaptive instance normalization:
    \begin{equation}
        \text{AdaIN}(x_i) = \gamma_i \frac{x_i - \mu(x_i)}{\sigma(x_i)} + \beta_i, 
    \end{equation}

where $x_i$ is an $i$-th channel of the input $x$, $(\gamma, \beta) = Aw+b$, A, b are learnable parameters, $w$ is found by propagating Gaussian noise through fully-connected network. This idea is taken from style-transfer papers \cite{style_transfer_real_time, artistic_style, stylization_network} and $\gamma$ is introduced as a \emph{style} variable. However for StyleGAN architecture this transformation is a part of random variables mapping and the key feature of such an approach is that it allows \emph{``modifying the relative importance of features for the subsequent convolution operation''} \cite{stylegan}
\end{itemize}

The discriminator of StyleGAN is a set of so-called downsampling blocks and is completed by a fully-connected, or \emph{dense}, layer. Downsampling block is depicted on Fig.~\ref{stylegan:discriminator}.
It is formed by one pooling and two convolutional layers. A dense layer is preceded by the \emph{Batch std} block (Fig.~\ref{discriminator:operations}). This block adds one more feature map to the output. The map is computed in the following way: for each component of the output in the batch standard deviations are computed, then the average is calculated, the average is replicated. Then it is added as a new channel to all the elements in the batch. 

\section{Generation of larger samples}\label{quilting}

StyleGAN architecture has fixed size input; this means that it can generate only samples of the size of the training data. However, from the practical point of view, it is more useful to be able to create images of different sizes. One way is to generate several samples and to locate them next to each other. However, this simple arrangement raises border artifacts. Another way is to modify the network architecture. We propose to generate several samples and then to use the procedure used for texture synthesis \,---\, image quilting \cite{quilting}.
The aim of this approach is to minimize the error between pixel values on the boundary between 2 images. The main steps are described in Algorithm \ref{quilting:algo}. This technique allows us to achieve smoother transitions between images.

\begin{algorithm}
\caption{\label{quilting:algo} Image quilting}

\SetAlgoLined
\KwInput{Two images $x$ and $y$ of size $N\times N$, overlap width $\omega$}
\KwOutput{Image $z$ of size $N\times 2N-\omega$}
\begin{enumerate}
    \item Place $x$ and $y$ next to each other with a overlap $\omega$ (Fig.~\ref{fig:quilting})
    \item Going along the overlap for each row $i = 1,\ldots,N$
    \begin{itemize}
        \item compute  $e_{i,j} = (x_{i,j}-y_{i,j})^2$
        \item find neighbouring pixel from the previous row $i-1$ with minimal resulting error on the border:\\
        $E_{i,j} = 
        \begin{cases} 
        e_{i,j},\, i = 1\\
        e_{i,j} + \min(E_{i-1,j-1}, E_{i-1,j}, E_{i-1, j+1}),\, \text{otherwise}
        \end{cases}$
        \item Remember the preceding pixel's location $l_{i,j}$ if $i\neq1$

    \end{itemize}
    
    \item $p \leftarrow \arg\min\limits_j E_{N, j}$
    \item for $i = N-1,\ldots, 1$:
    \begin{itemize}
        \item $z[:N-\omega+p] \leftarrow x[:N-\omega+p],\, z[N-\omega+p:] \leftarrow y[p:]$
        \item if $i\neq 1:$ $p = l_{i, p}$ 
    \end{itemize}

\end{enumerate}
\end{algorithm}

\begin{figure}[h]
    \centering
    \begin{tikzpicture}
    \node[text centered](img1) at (1,1) {Image 1};
    \path (img1.center) + (+2.4,0) node(img2) {Image 2};
    \begin{scope}
    \clip (2.5,1.8) -- (1.9, 1.8) -- (1.9, 0.2) -- (2.5,0.2) --cycle;
    \draw  [fill=green!20, decorate,decoration = {snake,amplitude =.4mm, segment length = 5mm}] (2.2,1.8)--(2.2,0.2) -- (2.2,-0.6) -- (1.5,-0.6) -- (1.5, 2) -- (2.2, 2) --cycle ;
    \end{scope}
    \begin{scope}
    \clip (2.5,1.8) -- (1.9, 1.8) -- (1.9, 0.2) -- (2.5,0.2) --cycle;
    \draw  [fill=orange!20, decorate,decoration = {snake,amplitude =.4mm, segment length = 5mm}] (2.2,1.8)--(2.2,0.2) -- (2.2,-0.6) -- (3,-0.6) -- (3, 2) -- (2.2, 2) --cycle ;
    \end{scope}
    \draw[orange!70] (1.9,1.8) -- (1.9, 0.2);
    \draw[green!70] (2.5,1.8) -- (2.5, 0.2);
    \draw [decorate,decoration={brace,amplitude=4pt, mirror}]
    (2.5, 1.8) -- (1.9, 1.8) node [black,midway,yshift=10pt] {$\omega$};
    \begin{pgfonlayer}{background}
            \path (img1.center)+(+1.5,+0.8) node (a) {};
            \path (img1.center)+(-1,-0.8) node (b) {};
            \path[draw=green!70]
                (a) rectangle (b);
                
            \path (img2.center)+(+1,+0.8) node (a) {};
            \path (img2.center)+(-1.5,-0.8) node (b) {};
            \path[draw=orange!70]
                (a) rectangle (b);
                
            \path (img2.center)+(+1,+0.8) node (a) {};
            \path (img2.center)+(-0.9,-0.8) node (b) {};
            \path[fill=orange!20]
                (a) rectangle (b);
            
            \path (img1.center)+(+0.9,+0.8) node (a) {};
            \path (img1.center)+(-1,-0.8) node (b) {};
            \path[fill=green!20]
                (a) rectangle (b);

    \end{pgfonlayer}
    \end{tikzpicture}
    \caption{Schematic result of image quilting}
    \label{fig:quilting}
\end{figure}
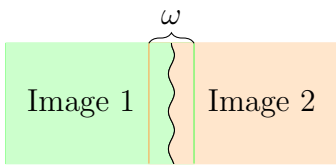

\section{Experiments}\label{experiments}

\subsection{Composites}\label{alporas_res}
\begin{figure*}
    \centering
    
    \begin{subfigure}[t]{0.3\textwidth}
    \includegraphics[width=0.9\textwidth]{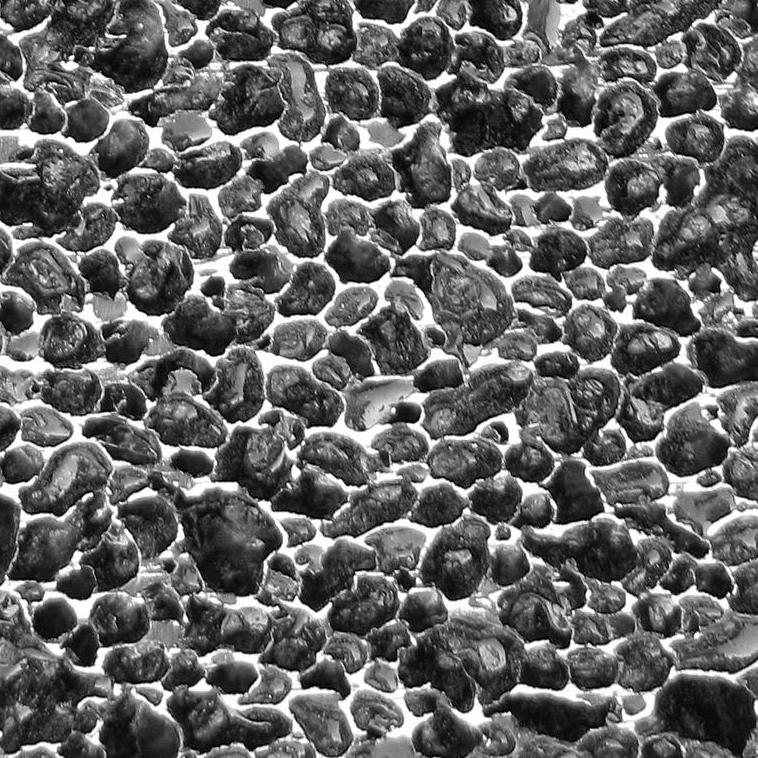}
    \caption{\label{alporas_res:original} Original image of alporas}
    \end{subfigure}
    \begin{subfigure}[t]{0.3\textwidth}
    \includegraphics[width=0.9\textwidth]{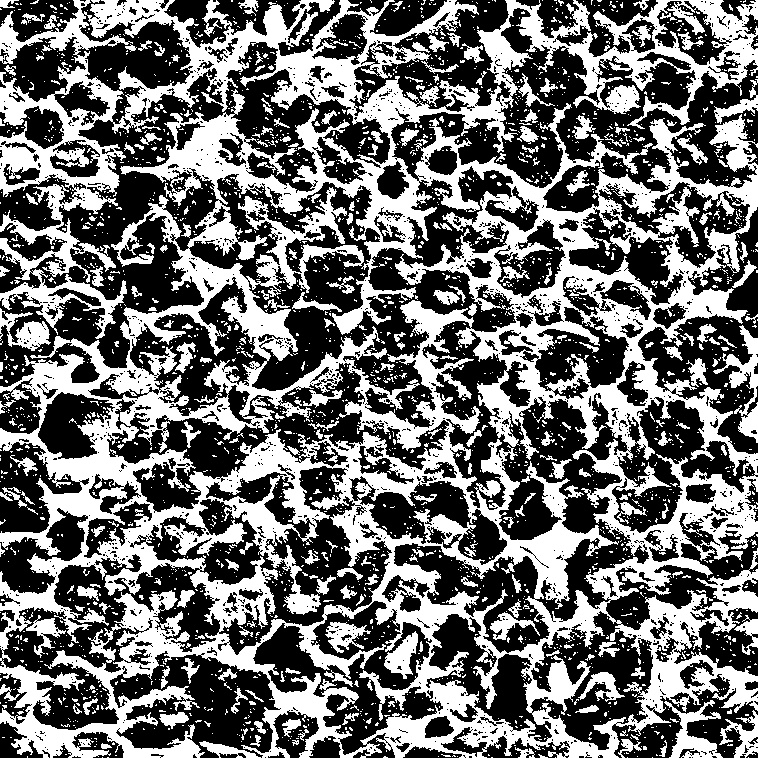}
    \caption{\label{alporas_res:binary}Binarized image with artifacts}
    \end{subfigure}
    \begin{subfigure}[t]{0.3\textwidth}
    \includegraphics[width=0.9\textwidth]{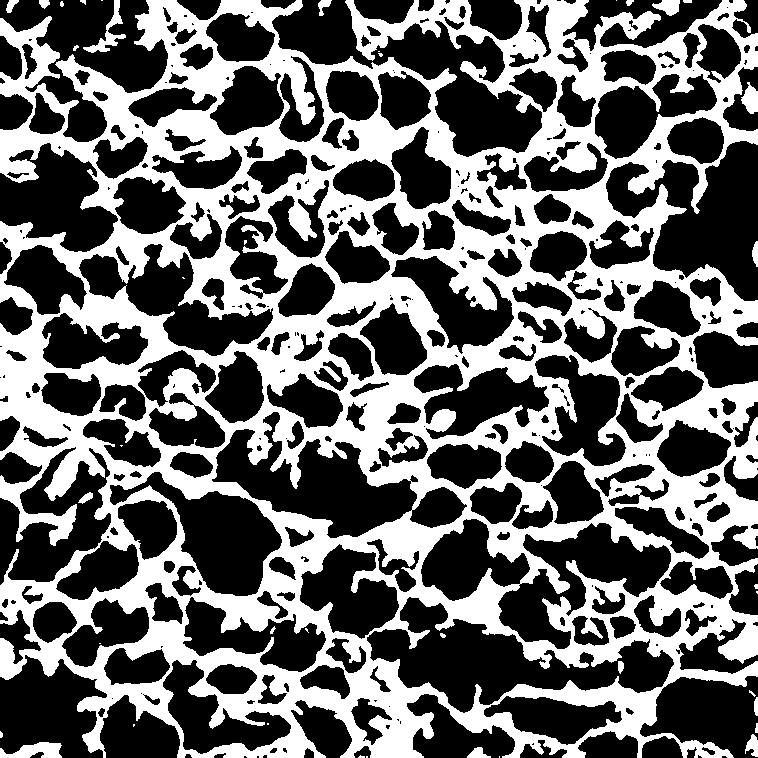}
    \caption{\label{alporas_res:filling_holes} Binarized image with filling of the holes}
    \end{subfigure}
    \begin{subfigure}[t]{0.3\textwidth}
    \includegraphics[width=0.9\textwidth]{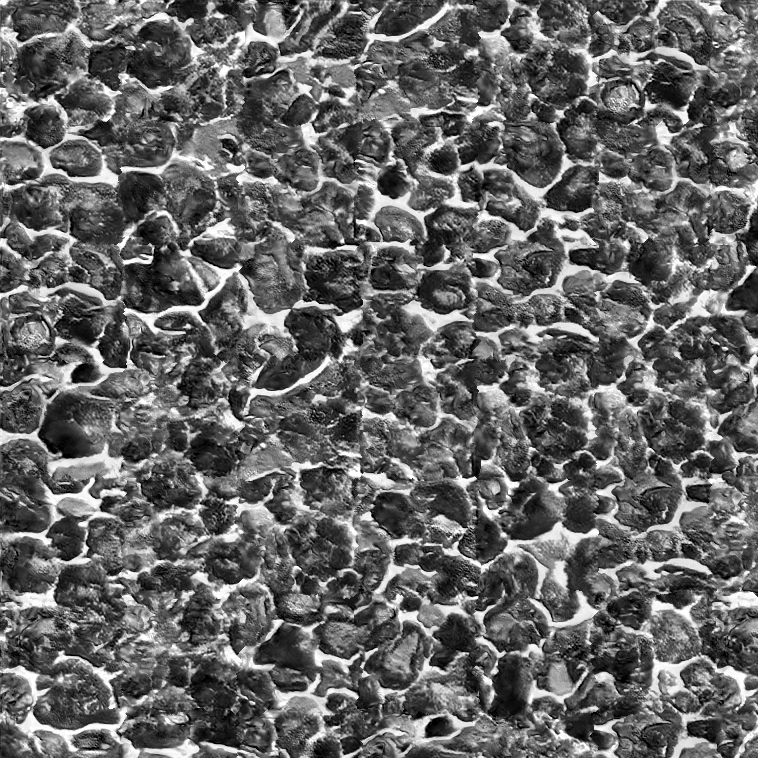}
    \caption{\label{alporas_res:stylegan} Result of StyleGAN}
    \end{subfigure}
    \begin{subfigure}[t]{0.3\textwidth}
    \includegraphics[width=0.9\textwidth]{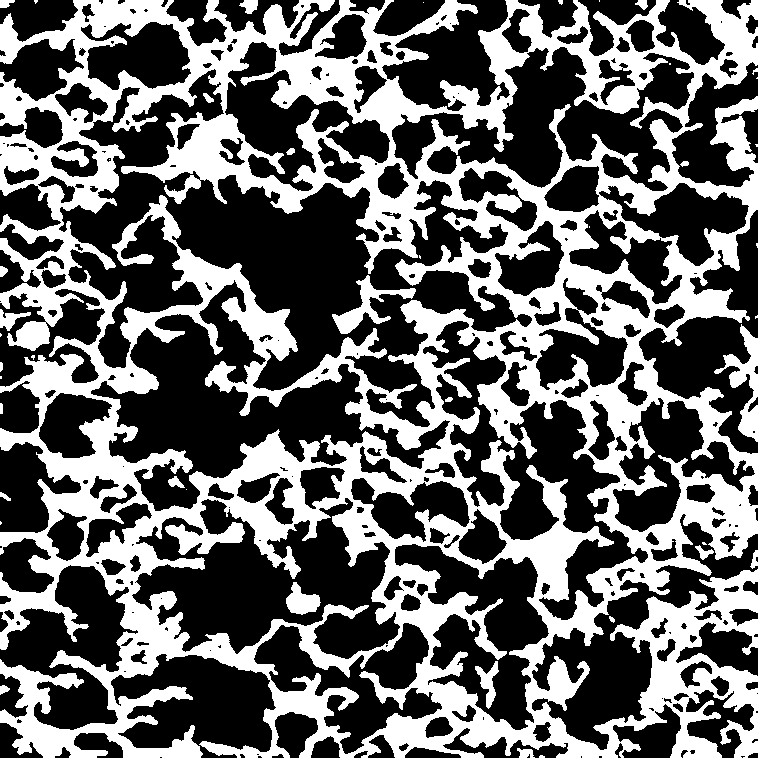}
    \caption{\label{alporas_res:stylegan_binary} Binarized StyleGAN result}
    \end{subfigure}
    
    \caption{\label{alporas_res:imgs} Visual comparison of obtained results for Alporas}
\end{figure*}

To test our method, we consider several structures. First example is a porous material \,---\, Alporas aluminum foam (Fig.~\ref{alporas_res:original}). This Alporas structure was previously studied in several works \cite{dovskavr2014aperiodic, metal_foams_characterization}. In the experiments we use an image by AlCarbon company \footnote{\url{https://www.stylepark.com/en/alcarbon/alporas-ac-black-both-sided-ground}}. This is a gray-scale image, where white areas correspond to the substance, gray areas correspond to the absence of the substance.

\begin{figure}[h]
    \centering
    \includegraphics[width=0.45\textwidth]{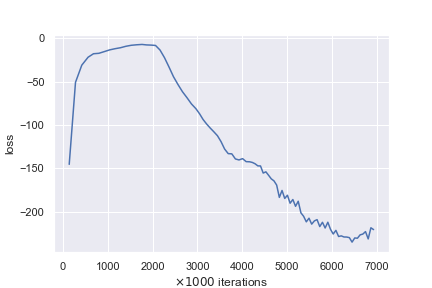}
    \caption{\label{alporas_res:loss}Training loss for Alporas images}
\end{figure}

\paragraph{Training.} 
We randomly cut the given $751\times751$ image into 16000 $128\times128$ samples and train the network on them. We use StyleGAN tensorflow implementation from GitHub repository \footnote{\url{https://github.com/NVlabs/stylegan}} and train with default parameters: Adam optimizer \cite{adam} with $\beta_1 = 0, \beta_2 = 0.99$ and learning rate \,---\, 0.001 (it is increased to 0.0015 when the size of synthesized images 128), batch size of 16. The size of training images gradually increases from $8\times8$ to $128\times128$. Convergence of the method is shown on Fig.~\ref{alporas_res:loss}.

\paragraph{Postprocessing and evaluation.}
To estimate the quality of reconstruction, we compute its effective elastic properties \,---\, Young's modulus ($E$) and Poisson's ratio ($\nu$). They are computed via homogenization procedure \cite{Panetta}. Fenics platform \cite{fenics} is used for numerical computations. The input of the algorithm is a binary image. However, the generated images are gray-scale. We manually find the threshold for image binarization (equals 116 for both original and synthesized structures). As the original image has light-gray areas in centers of pores (Fig.~\ref{alporas_res:original}), the thresholding produces artifacts in these centers (Fig.~\ref{alporas_res:binary}). With a larger threshold, we partly remove the substance. For our example it is quite easy to eliminate these artifacts. We fill the holes via scipy.ndimage Python package, then use Gaussian filter \cite{image_processing} with the kernel of size (5,5) and standard deviation 20 to smooth the edges, and finally binarize the result with Otsu threshold \cite{otsu1979threshold}.

\paragraph{Results.}
Table~\ref{alporas_res:mech_properties} shows the numerical evaluation of the obtained structures. The presented values are the estimated mean and variance of Young's modulus and Poisson's ratio. For this purpose, 50 samples were used in case of original and in case of synthesized images.

\begin{table}[h]
    \caption{\label{alporas_res:mech_properties_1} Elastic properties for one original and one generated samples}
    \begin{ruledtabular}
    \begin{tabular}{>{\centering\arraybackslash}m{.3\linewidth}>{\centering\arraybackslash}m{.3\linewidth}>{\centering\arraybackslash}m{.3\linewidth}}
         & $E$ & $\nu$\\
        \hline
         Original image & $0.1275$ & $0.3703$ \\
        Synthesized image & $0.1175$ & $0.3472$ \\
    \end{tabular}
    \end{ruledtabular}

\end{table}

\begin{table}[h]
    \caption{\label{alporas_res:mech_properties} Elastic properties of original and generated samples}
    \begin{ruledtabular}
     \begin{tabular}{>{\centering\arraybackslash}m{.3\linewidth}>{\centering\arraybackslash}m{.3\linewidth}>{\centering\arraybackslash}m{.3\linewidth}}
         & $E$ & $\nu$\\
        \hline
        Original image & $0.0998\pm 0.0067$ & $0.3716\pm 0.0119$ \\
        Synthesized image & $0.1002\pm 0.0122$ & $0.3677\pm 0.0335$ \\
    \end{tabular}
    \end{ruledtabular}
\end{table}

\subsection{Digital Rock}\label{dig_rock}

Another important application area is the analysis of tomography images. Recent studies have applied GAN models to this data, and we can compare out generation results with theirs. We took two benchmark three-dimensional micro-CT images from Imperial College London collection \footnote{\url{https://www.imperial.ac.uk/earth-science/research/research-groups/perm/research/pore-scale-modelling/micro-ct-images-and-networks/}}. They are images of Berea sandstone and Ketton limestone. For training on Berea 10000 two-dimensional slices were randomly cut from them, for Ketton - 10240. 

\begin{figure*}
    \begin{subfigure}{0.45\textwidth}
    \centering
    \includegraphics[width=\textwidth]{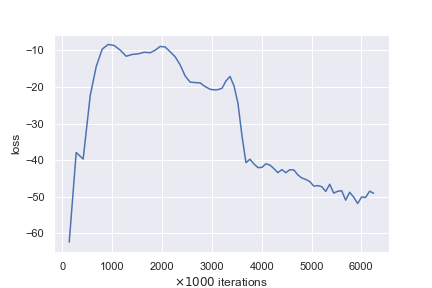}
    \caption{Berea dataset}
    \label{berea_res:loss}
    \end{subfigure}
    \begin{subfigure}{0.45\textwidth}
    \centering
    \includegraphics[width=\textwidth]{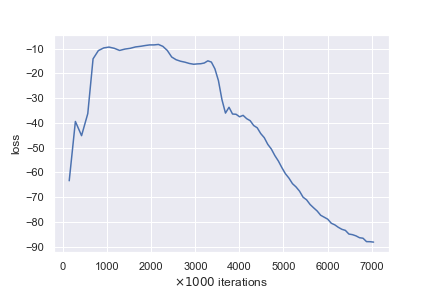}
    \caption{Ketton dataset}
    \label{ketton_res:loss}
    \end{subfigure}
    \caption{\label{dig_rock:losses}Training losses}
\end{figure*}

\paragraph{Training.}
We use the same StyleGAN tensorflow implementation \cite{stylegan} and train with default parameters: Adam optimizer with $\beta_1 = 0, \beta_2 = 0.99$ and learning rate \,---\, 0.001, batch size of 16.  The size of training images gradually increases from $8\times8$ to $64\times64$. Convergence of the method is shown on Fig.~\ref{dig_rock:losses}.

\paragraph{Postprocessing and evaluation.}
The original images are binary. We smooth the gray-scale image with median filter \cite{image_processing} and then binarize via Otsu method \cite{otsu1979threshold}. We use Minkowski functionals \cite{minkowski} to compare the generated and real structures. For a 2D structure, there are three of them:
\begin{enumerate}
    \item Area density:
    $\frac{S_{\text{voids}}}{S}$, where $S$ - area of sample, $S_{\text{voids}}$ - area of the voids
    \item Perimeter density:
    $\frac{L}{S}$, where $L$ - perimeter of grains
    \item Euler characteristic:
    $\chi = \frac{V-E+F}{S},$ where $V$ - number of vertices, $E$ - number of edges, $F$ - number of regions
\end{enumerate}

\begin{figure*}[h]
    \centering
    \begin{subfigure}[t]{0.24\textwidth}
    \includegraphics[width=0.8\textwidth, frame]{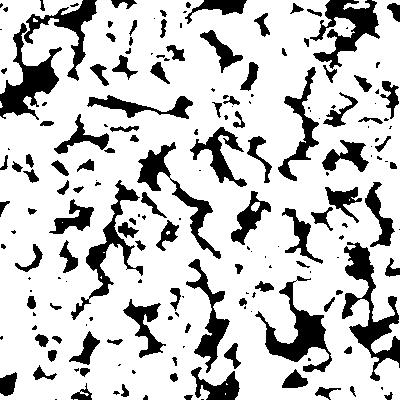}
    \caption{Original image}
    \label{dig_rock:berea_original}

    \end{subfigure}
    \begin{subfigure}[t]{0.24\textwidth}
    \includegraphics[width=0.8\textwidth, frame]{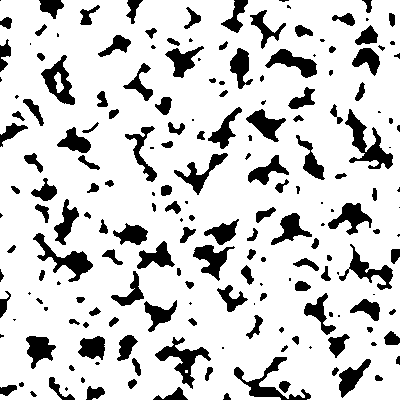}

    \caption{Result of StyleGAN + quilting}
    \label{dig_rock:berea_stylegan}
    \end{subfigure}
    \begin{subfigure}[t]{0.24\textwidth}
    \includegraphics[width=0.8\textwidth, frame]{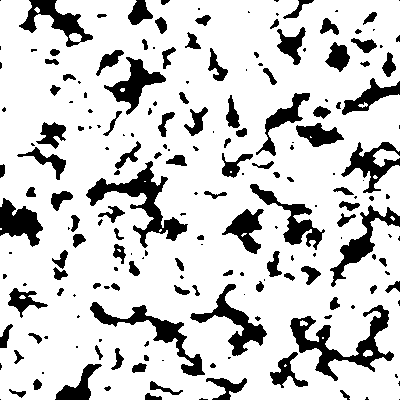}
    \caption{Result of PorousMediaGAN}
    \label{dig_rock:berea_pormediagan}

    \end{subfigure}
    \caption{Visual comparison of the results for Berea}
    \label{dig_rock:berea}
\end{figure*}

\begin{figure*}[h]
    \centering
    \begin{subfigure}[t]{0.24\textwidth}
    \includegraphics[width=0.8\textwidth, frame]{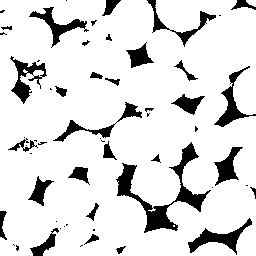}
    \caption{Original image}
    \label{dig_rock:ketton_original}
    \end{subfigure}
    \begin{subfigure}[t]{0.24\textwidth}
    \includegraphics[width=0.8\textwidth, frame]{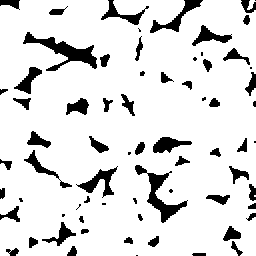}
    \caption{Result of StyleGAN + quilting}
    \label{dig_rock:ketton_stylegan}
    \end{subfigure}
    \begin{subfigure}[t]{0.24\textwidth}
    \includegraphics[width=0.8\textwidth, frame]{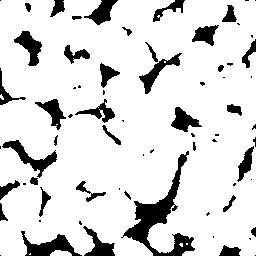}
    \caption{Result of PorousMediaGAN}
    \label{dig_rock:ketton_pormediagan}
    \end{subfigure}
    \caption{Visual comparison of the results for Ketton}
    \label{dig_rock:ketton}
\end{figure*}

\paragraph{Results.}
Figures \ref{dig_rock:berea_stylegan}, \ref{dig_rock:ketton_stylegan} display the resulting structures. For Berea the size of image is $400\times400$, for Ketton - $256\times256$. These structures were compared with images of the original structure, as well as samples, produced by Porous Media GAN \cite{mosser}. In case of Porous Media GAN, for comparison, we take two-dimensional slices of a three-dimensional image synthesized by the pretrained model for comparison. Corresponding values of Minkowski functionals are presented in Tables~\ref{dig_rock:minkowski_berea_1}, \ref{dig_rock:minkowski_ketton_1}.

\begin{table}[!h]
\caption{\label{dig_rock:minkowski_berea_1} Comparison of Minkowski functionals for $400\times400$ samples of Berea}
\begin{ruledtabular}
\begin{tabular}{>{\centering\arraybackslash}m{.2\linewidth}>{\centering\arraybackslash}m{.2\linewidth}>{\centering\arraybackslash}m{.2\linewidth}>{\centering\arraybackslash}m{.2\linewidth}}
         & Area & Perimeter & Euler2d\\
        \hline
        Original& $0.7810$ & $0.0724$ & $-0.0010$  \\
        PorousMediaGAN & $0.8072$ & $0.0625$ & $-0.0010$\\
        StyleGAN & $0.8174$ & $0.0605$ & $-0.0011$ \\
    \end{tabular}
\end{ruledtabular}

\end{table}

\begin{table}[!h]
\caption{\label{dig_rock:minkowski_ketton_1} Comparison of Minkowski functionals for $256\times256$ samples of Ketton}
\begin{ruledtabular}
\begin{tabular}{>{\centering\arraybackslash}m{.2\linewidth}>{\centering\arraybackslash}m{.2\linewidth}>{\centering\arraybackslash}m{.2\linewidth}>{\centering\arraybackslash}m{.2\linewidth}}
         & Area & Perimeter & Euler2d\\
        \hline
        Original & $0.8687$ & $0.0513$ & $-0.0007$  \\
        PorousMediaGAN & $0.8586$ & $0.0482$ & $-0.0008$\\
        StyleGAN & $0.8749$ & $0.0507$ & $-0.0009$ \\
    \end{tabular}
\end{ruledtabular}
\end{table}

\begin{figure*}
    \centering
    \begin{subfigure}{0.45\textwidth}
    \includegraphics[width=\textwidth]{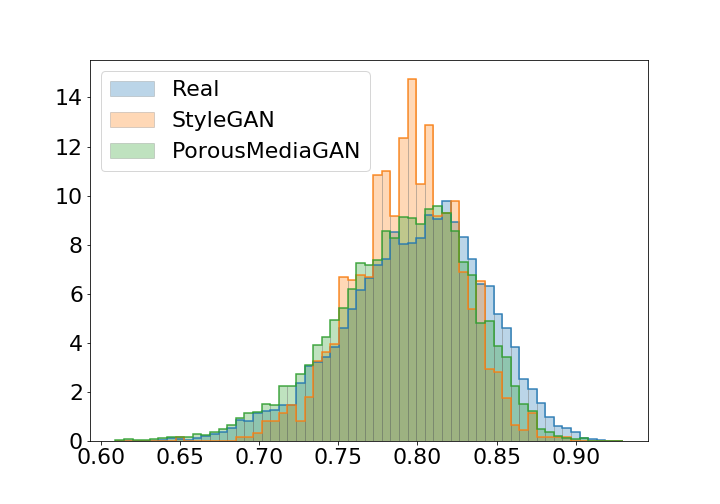}
    \caption{Distribution of area density}
    \end{subfigure}
    \begin{subfigure}{0.45\textwidth}
    \includegraphics[width=\textwidth]{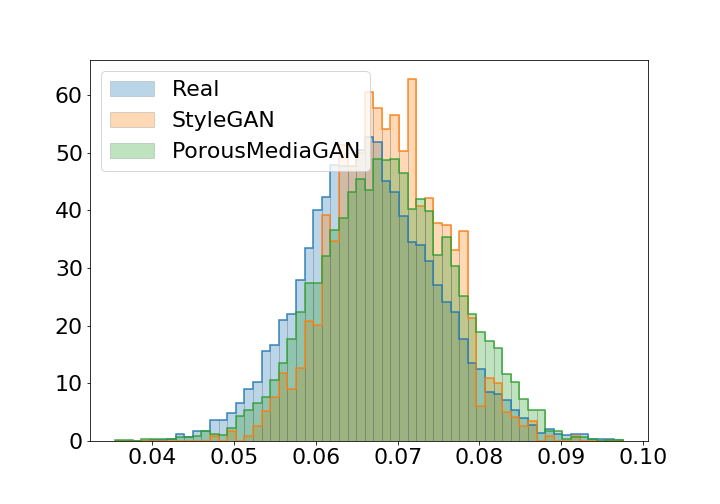}
    \caption{Distribution of perimeter density}
    \end{subfigure}
    \begin{subfigure}{0.45\textwidth}
    \includegraphics[width=\textwidth]{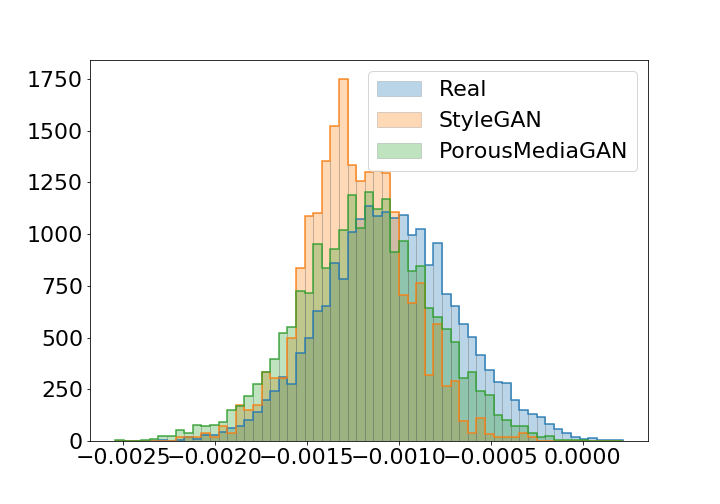}
    \caption{Distribution of Euler2d density}
    \end{subfigure}
    
    \caption{Minkowski functionals distributions for Berea}
    \label{dig_rock:minkowski_hist_berea}
\end{figure*}

\begin{figure*}
    \centering
    \begin{subfigure}{0.45\textwidth}
    \includegraphics[width=\textwidth]{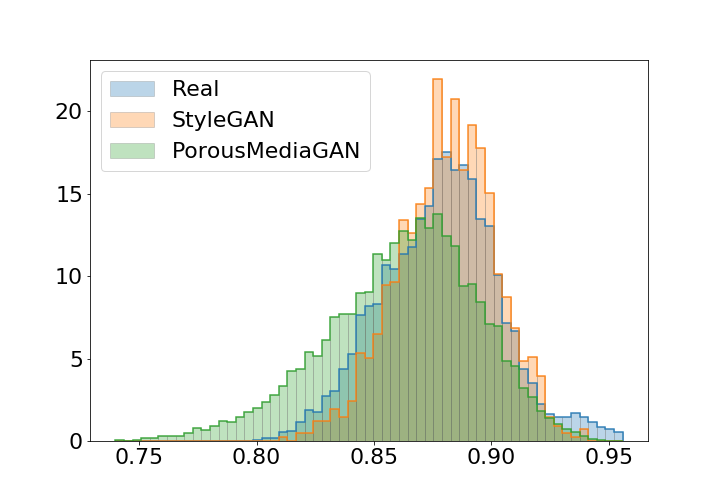}
    \caption{Distribution of area density}
    \end{subfigure}
    \begin{subfigure}{0.45\textwidth}
    \includegraphics[width=\textwidth]{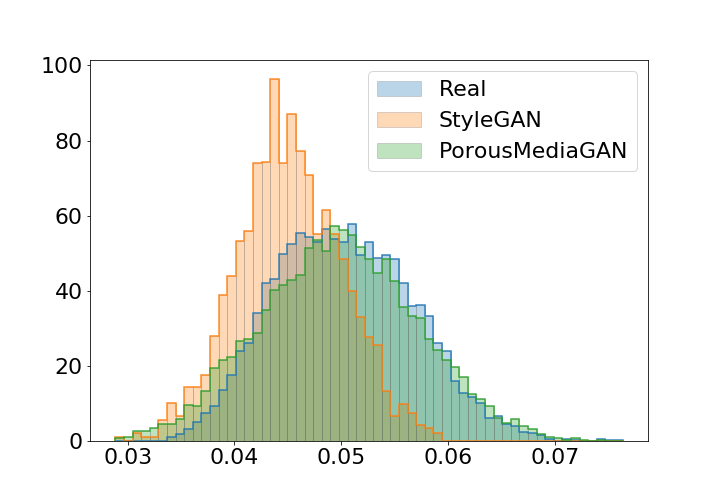}
    \caption{Distribution of perimeter density}
    \end{subfigure}
    \begin{subfigure}{0.45\textwidth}
    \includegraphics[width=\textwidth]{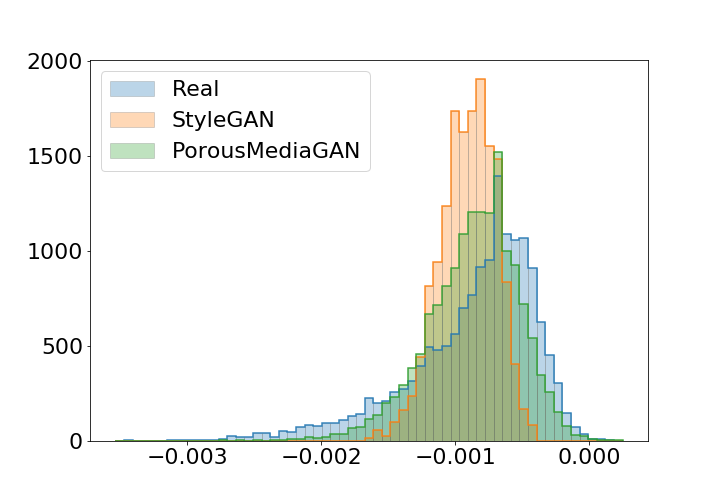}
    \caption{Distribution of Euler2d density}
    \end{subfigure}
    
    \caption{Minkowski functionals distributions for Ketton}
    \label{dig_rock:minkowski_hist_ketton}
\end{figure*}

Tables~\ref{dig_rock:minkowski_berea}, \ref{dig_rock:minkowski_ketton} show average values of Minkowski functionals for three types of structures: original, synthesised by PorousMediaGan \cite{mosser} and generated by the new StyleGAN approach. The empirical distributions of these values can be viewed on Fig.~\ref{dig_rock:minkowski_hist_berea}, \ref{dig_rock:minkowski_hist_ketton}. The size of the samples used for comparison is 128. 
\begin{table*}
    \caption{\label{dig_rock:minkowski_berea}Comparison of Minkowski functionals for Berea}
    \begin{ruledtabular}
    \begin{tabular}{>{\centering\arraybackslash}m{.2\linewidth}>{\centering\arraybackslash}m{.2\linewidth}>{\centering\arraybackslash}m{.2\linewidth}>{\centering\arraybackslash}m{.2\linewidth}}
         & Area & Perimeter & Euler2d\\
        \hline
        Original & $0.7984\pm 0.0439$ & $0.0666\pm 0.0082$ & $-0.0010 \pm 0.0004$  \\
        PorousMediaGAN & $0.7900\pm 0.0433$ & $0.0689\pm 0.0083$ & $-0.0012\pm 0.0004$\\
        StyleGAN & $0.7934\pm 0.0334$ & $0.0690\pm 0.0066$ & $-0.0012\pm 0.0003$ \\
    \end{tabular}
    \end{ruledtabular}
\end{table*}

\begin{table*}
\caption{\label{dig_rock:minkowski_ketton} Comparison of Minkowski functionals for Ketton}
\begin{ruledtabular}
    \begin{tabular}{>{\centering\arraybackslash}m{.2\linewidth}>{\centering\arraybackslash}m{.2\linewidth}>{\centering\arraybackslash}m{.2\linewidth}>{\centering\arraybackslash}m{.2\linewidth}}
         & Area & Perimeter & Euler2d\\
        \hline
        Original & $0.8783\pm 0.0252$ & $0.0503\pm 0.0065$ & $-0.0009 \pm 0.0004$  \\
        PorousMediaGAN & $0.8623\pm 0.0321$ & $0.0498\pm 0.0073$ & $-0.0009\pm 0.0003$\\
        StyleGAN & $0.8810\pm 0.0208$ & $0.0452\pm 0.0050$ & $-0.0009\pm 0.0002$ \\
    \end{tabular}
\end{ruledtabular}
\end{table*}

\subsection{Comparison of a multi-resolution GAN and a GAN, using only one resolution}\label{np_pg}
To define if we need the gradual resolution growth, we compare the results of the training with the sequential adding of the blocks and results of the direct training of the full network. We estimate mean and variance of the considered properties in the same way as in Section~\ref{experiments} and present in \Crefrange{np_pg:mech_properties}{np_pg:minkowski_ketton}. The mean values are quite close. One can also notice that the difference between the usage of progressive growing and training without it is not significant. Though, on average training with different resolutions has a better result.

\begin{table*}[]
    \caption{\label{np_pg:mech_properties} Elastic properties of original and generated samples for two types of training ($128\times128$ samples)}
    \begin{ruledtabular}
    \begin{tabular}{>{\centering\arraybackslash}m{.47\linewidth}>{\centering\arraybackslash}m{.24\linewidth}>{\centering\arraybackslash}m{.24\linewidth}}
         & $E$ & $\nu$ \\
         \hline
        Original &  $0.1019 \pm 0.0116$ & $0.3660 \pm 0.0385$ \\
        With progressive growing & $0.1093 \pm 0.0161$ & $0.3670 \pm 0.0316$ \\
        Without progressive growing & $0.1075 \pm 0.0132$ & $0.3698 \pm 0.0324$ \\
    \end{tabular}
    \end{ruledtabular}
\end{table*}

\begin{table*}[]
    \caption{\label{np_pg:minkowski_berea} Minkowski functionals of original and generated samples for two types of training for Berea ($64\times64$ samples)}
    \begin{ruledtabular}
    \begin{tabular}{>{\centering\arraybackslash}m{.25\linewidth}>{\centering\arraybackslash}m{.25\linewidth}>{\centering\arraybackslash}m{.25\linewidth}>{\centering\arraybackslash}m{.25\linewidth}}
         & Area & Perimeter & Euler2D \\
        \hline
        Original &  $0.7987 \pm 0.0800$ & $0.0666 \pm 0.0159$ & $-0.0011 \pm 0.0007$ \\
        With progressive growing & $0.7972 \pm 0.0531$ & $0.0644 \pm 0.0109$ & $-0.0011 \pm 0.0005$ \\
        Without progressive growing & $0.8055 \pm 0.0108$ & $0.0616 \pm 0.0089$ & $-0.0012 \pm 0.0004$ \\
    \end{tabular}
    \end{ruledtabular}
\end{table*}
\begin{table*}[]
    \caption{\label{np_pg:minkowski_ketton} Minkowski functionals of original and generated samples for two types of training for Ketton ($64\times64$ samples)}
    \begin{ruledtabular}
    \begin{tabular}{>{\centering\arraybackslash}m{.25\linewidth}>{\centering\arraybackslash}m{.25\linewidth}>{\centering\arraybackslash}m{.25\linewidth}>{\centering\arraybackslash}m{.25\linewidth}}
         & Area & Perimeter & Euler2D \\
        \hline
        Original &  $0.8780 \pm 0.0484$ & $0.0495 \pm 0.0131$ & $-0.0008 \pm 0.0007$ \\
        With progressive growing & $0.8760 \pm 0.0303$ & $0.0445 \pm 0.0067$ & $-0.0009 \pm 0.0004$ \\
        Without progressive growing & $0.8802 \pm 0.0292$ & $0.0430 \pm 0.0073$ & $-0.0009 \pm 0.0004$ \\
    \end{tabular}
    \end{ruledtabular}
\end{table*}

\section{Increasing the resolution of samples.}\label{res_increase}
We may also alter the StyleGAN architecture so that it will be able to increase an image resolution. We force the generator to produce a two-times bigger image by adding one more upsampling block, then take every second pixel of the generated image along each axis and forward to the discriminator. For this approach we examine the necessity of progressive growing and estimate elastic properties for Alporas samples and Minkowski functionals for digital rock data as well. All samples have the same resolution \,---\, $128\times128$. As it can be seen from \Crefrange{res_increase:mech_properties}{res_increase:minkowski_ketton}, the difference between the considered values is small. However, for Berea data (Table~\ref{res_increase:minkowski_berea}) the usage of progressive growing allows obtaining a larger variance, which is close to the variance of real data.

\begin{table*}[]
    \caption{\label{res_increase:mech_properties} Elastic properties of original and generated samples for two types of training for Alporas (increasing the resolution)}
    \begin{ruledtabular}
    \begin{tabular}{>{\centering\arraybackslash}m{.47\linewidth}>{\centering\arraybackslash}m{.24\linewidth}>{\centering\arraybackslash}m{.24\linewidth}}
         & $E$ & $\nu$ \\
        \hline
        Original &  $0.1019 \pm 0.0116$ & $0.3660 \pm 0.0385$ \\
        With progressive growing & $0.1047 \pm 0.0171$ & $0.3645 \pm 0.0390$ \\
        Without progressive growing & $0.1084 \pm 0.0166$ & $0.3631 \pm 0.0400$ \\
    \end{tabular}
    \end{ruledtabular}
\end{table*}

\begin{table*}[]
    \caption{\label{res_increase:minkowski_berea} Minkowski functionals of original and generated samples for two types of training for Berea (increasing the resolution)}
    \begin{ruledtabular}
    \begin{tabular}{>{\centering\arraybackslash}m{.25\linewidth}>{\centering\arraybackslash}m{.25\linewidth}>{\centering\arraybackslash}m{.25\linewidth}>{\centering\arraybackslash}m{.25\linewidth}}
         & Area & Perimeter & Euler2D \\
        \hline
        Original & $0.7984 \pm 0.0439$ & $0.0666 \pm 0.0082$ & $-0.0010 \pm 0.0004$  \\
        With progressive growing & $0.8016 \pm 0.0385$  & $0.0646 \pm 0.0052$ & $-0.0010 \pm 0.0003$ \\
        Without progressive growing & $0.7987 \pm 0.0061$ & $0.0615 \pm 0.0049$ & $-0.0009 \pm 0.0002$ \\
    \end{tabular}
    \end{ruledtabular}
\end{table*}
\begin{table*}[]
    \caption{\label{res_increase:minkowski_ketton} Minkowski functionals of original and generated samples for two types of training for Ketton (increasing the resolution)}
    \begin{ruledtabular}
    \begin{tabular}{>{\centering\arraybackslash}m{.25\linewidth}>{\centering\arraybackslash}m{.25\linewidth}>{\centering\arraybackslash}m{.25\linewidth}>{\centering\arraybackslash}m{.25\linewidth}}
         & Area & Perimeter & Euler2D \\
        \hline
        Original & $0.8776\pm 0.0254$ & $0.0503\pm 0.0065$ & $-0.0009 \pm 0.0005$  \\
        With progressive growing & $0.8698 \pm 0.0167$ & $0.0464 \pm 0.0047$ & $-0.0008 \pm 0.0003$ \\
        Without progressive growing & $0.8715 \pm 0.0195$ & $0.0560 \pm 0.0049$ & $-0.0009 \pm 0.0003$ \\
    \end{tabular}
    \end{ruledtabular}
\end{table*}

\section{Conclusion and future work}\label{conclusion}
\begin{figure}
    \centering
    \scalebox{0.9}{
    \begin{tikzpicture}
    \node[inner sep=0pt] (train) at (0,0)
    {\includegraphics[width=.2\textwidth, frame]{images/ketton_256.png}};
    \node[inner sep=0pt] (cuts) at (+5.5,0)
    {\includegraphics[width=.2\textwidth]{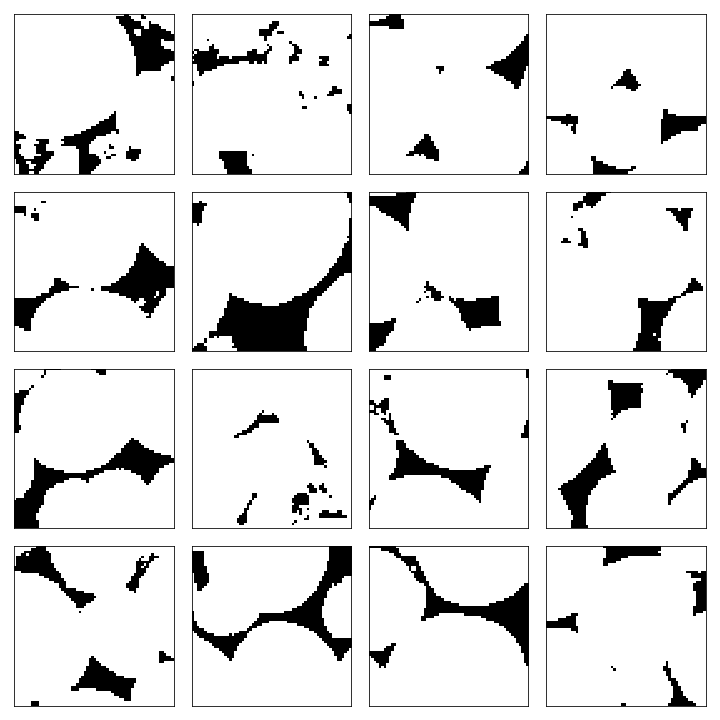}};
    \node[inner sep=0pt] (cuts_gen) at (5.5,-5.5)
    {\includegraphics[width=.2\textwidth]{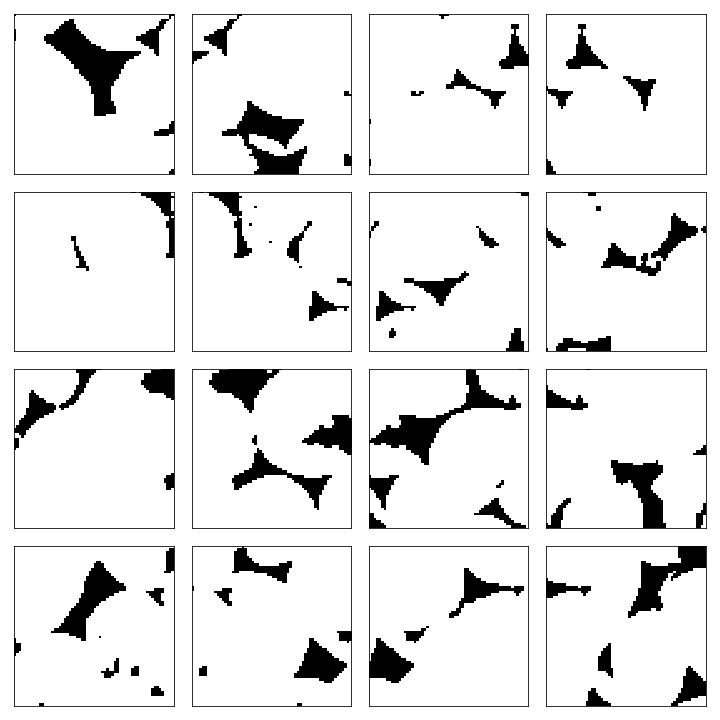}};
    \node[inner sep=0pt] (res) at (0,-5.5)
    {\includegraphics[width=.2\textwidth, frame]{images/ketton_final_256.png}};
    
    \path (train.center)+(+2,0) node(node1){};
    \path (train.center)+(+3.5,0) node(node2){};
    \path (train.center)+(+2.8,+0.4) node (cut) {cut};
    \path (train.center)+(+2.8,-0.4) node (cut1) {image};
    \draw[-latex,thick] (node1.center) -- (node2.center);
    
    \path (cuts.center)+(0,-2) node(node3){};
    \path (cuts.center)+(0,-3.5) node(node4){};
    \path (cuts.center)+(+0.9,-2.3) node (gen) {generate};
    \path (cuts.center)+(+0.9,-2.7) node (gen) {new};
    \path (cuts.center)+(+0.9,-3.1) node (gen) {samples};
    \draw[-latex,thick] (node3.center) -- (node4.center);
    
    \path (res.center)+(+2,0) node(node1){};
    \path (res.center)+(+3.5,0) node(node2){};
    \path (res.center)+(+2.8,+0.4) node (cut) {merge};
    \path (res.center)+(+2.8,-0.4) node (cut1) {samples};
    \draw[-latex,thick] (node2.center) -- (node1.center);
    
    \end{tikzpicture}
    }
    \caption{The process of generating new structures}
    \label{conclusion:pipeline}
\end{figure}

The overall pipeline of work is shown on Fig.~\ref{conclusion:pipeline}. 
The proposed method has shown its efficiency for the considered task. Empirical distributions of properties of samples produced by the style-based network (Figs.~\ref{dig_rock:minkowski_hist_berea}, \ref{dig_rock:minkowski_hist_ketton}) are very similar to the distributions for original samples. The presented numerical values display a better result of StyleGAN on average in comparison with Porous Media GAN. This result highlights the high capability of this architecture to embed information about the microstructure and then reproduce it. Moreover, the progressive growing increases the variation of the produced samples.

However, this approach requires quite a tricky postprocessing procedure with the manual selection of the parameters. Our method includes filling the holes, to be exact, white areas that are not connected to the borders. The bigger the images are, the worse the result is, as larger generated samples have more isolated white areas. As one can see on Figs.~\ref{dig_rock:minkowski_hist_berea}, \ref{dig_rock:minkowski_hist_ketton}, histograms for StyleGAN samples have high peaks. Also, though for area density the distribution is close to the real, we can observe that perimeter and Euler characteristic distributions are displaced in comparison with original images. The displacement is caused by the usage of image quilting. This means that a better method for increasing samples' sizes needs to be found.

For our approach another extension can be developed. In our work we consider 2D images. However, from the physical point of view, it is more practical to have a three-dimensional representation of the structure. To our knowledge, there were no attempts to apply the style-based GAN architecture for three-dimensional images. Taking into account remarkable results for the two-dimensional case, we propose to extend our method to the 3D case. 

\FloatBarrier
\bibliography{apssamp}

\end{document}